\documentclass[prd,aps,nofootinbib,preprint]{revtex4}
\usepackage{amsmath,amssymb,amsfonts,color,graphicx,graphics,latexsym,placeins,epsfig,multirow}
\usepackage{array}
\newcolumntype{P}[1]{>{\centering\arraybackslash}p{#1}}
\usepackage{footmisc}
\usepackage[compatibility=false]{caption}
\usepackage{subcaption}
\usepackage{bbold}
\usepackage{enumitem}
\setdescription{leftmargin=12.5pt}
\usepackage{bbold}
\usepackage{hyperref}
\DeclareMathOperator{\sech}{sech}
\usepackage{natbib}
\setcitestyle{square, comma, numbers, sort&compress}
\usepackage{float}
\newcommand*{\id}{{\rm\hbox{1\kern-0.15em \vrule width .1pt depth-.2pt}}}

\begin{document}


\title{\Large \bf Kundt geometries and memory effects in the Brans-Dicke theory of gravity}
\author{Siddhant Siddhant, Indranil Chakraborty and
Sayan Kar}
\email{siddhant@iitkgp.ac.in, indradeb@iitkgp.ac.in,   sayan@phy.iitkgp.ac.in}
\affiliation{Department of Physics \\ Indian Institute of Technology Kharagpur, 721 302, India}

\begin{abstract}
 \noindent {Memory effects 
 are studied in 
 the simplest scalar-tensor theory, the Brans--Dicke (BD) theory.  To this end, we introduce, in
 BD theory, novel Kundt spacetimes (without and with
 gyratonic terms), which
 serve as backgrounds for the ensuing analysis on memory. The BD parameter $\omega$ and the scalar field ($\phi$) profile, expectedly, distinguishes between different solutions.
 Choosing specific localised forms for the
 free metric functions $H'(u)$ (related to the wave profile)
 and $J(u)$ (the gyraton) we obtain displacement memory
 effects using both geodesics and geodesic deviation. 
 An interesting and easy-to-understand
 exactly solvable case arises when $\omega=-2$ (with
 $J(u)$ absent) which we discuss in detail. For other $\omega$ (in the presence of
 $J$ or without), numerically
 obtained geodesics lead to results on
 displacement memory which appear to match
 qualitatively with those found from a deviation analysis. 
 Thus, the issue of how 
 memory effects in BD theory may arise and also
 differ from their GR counterparts, is now 
 partially addressed, at least theoretically, within the
 context of this new class of Kundt geometries.
 }

\end{abstract}

\maketitle

\section{Introduction}

\noindent The detection of gravitational waves in binary mergers has opened up new prospects for testing theories of gravity in the strong field regime \citep{Abbott:2016,Berti:2018}. Gravitational wave memory is one such as-yet-unobserved strong field effect that can be used to test diverse theories of gravity. The gravitational wave memory effect is the residual permanent shift in the position (or velocity) caused due to the passage of a gravitational wave pulse \citep{Favata:2010}.

\noindent The study of memory effects began in the work of Zel'dovich and Polnarev \citep{Zeldovich:1974} who studied gravitational radiation emitted due to the motion of flybys/collapse of stars in a globular cluster. A few years later, Braginsky and Grishchuk \citep{Braginsky:1985} 
looked at the deviation of test particles in weak field, linearized gravity and coined the term {\em memory effect} to denote the change in the metric perturbation at early and late times. Geodesic deviation of test particles due to low frequency gravitational radiation at null infinity was investigated further by Ludvigsen \citep{Ludvigsen:1989}.  Christodoulou, using full nonlinear GR, attributed the presence of memory to the transport of energy and momentum of gravitational waves 
to null infinity \citep{Christodoulou:1991}. Further, this effect, related to non-linearity, was ascribed to gravitons produced by the radiation itself \citep{,Thorne:1991}. Memory effects are also possible in electrodynamics \citep{Bieri:2013} and Yang-Mills theories \citep{Pate:2017,Jokela:2019}. Interesting theoretical links to memory effects have been conjectured, of late, in the context of soft theorems and BMS symmetries \citep{Strominger:2016}. It has been noted that the nonlinear memory effect can also be understood as a BMS transformation relating two inequivalent Minkowski vacua at future null infinity caused by the passage of gravitational waves (see the review \citep{Strominger:2017} and the references cited therein).

 \noindent
Apart from spacetime boundaries, memory effects can also be realized in the interior regions of a spacetime \citep{Zhang:2017,Chak:2020}. Such effects lead to permanent changes due to presence of gravitational waves and have been termed as {\em persistent observables} in \citep{Flanagan:2019}. Our work mainly focuses on one such observable named {\em displacement memory observable} and tries to calculate the memory in non-flat Kundt geometries.

\noindent Memory effects in non-flat backgrounds in GR have been studied in both dS \citep{Hamada:2017,Bieri:2017} (motivations from cosmology) and AdS spacetimes \citep{Chu:2019}. In \citep{Chu:2019}, the authors have showed how to isolate the gravitational wave contribution from the background spacetime by resorting to Fermi normal coordinates and solving the geodesic deviation equation. They treated the wave as a perturbation over AdS spacetime caused due to scattering of massive/massless particles. In our work, we adopt the same method for studying memory effects in Kundt spacetimes. However, in our case, the setting is non-perturbative, since we deal with
exact spacetimes representing gravitational waves.

\noindent Kundt spacetimes are exact radiative geometries  consisting of non-expanding, non-shearing and non-twisting null geodesic congruences (NGC) \citep{Kundt:1961,Stephani:2003,Griffiths:2009}. They admit various wave solutions ({\em pp} waves, Siklos waves \citep{Brinkmann:1925,Rosen:1937}) 
related to the presence of NGC whose tangent vector is usually not covariantly constant. In general, the wave surfaces may not be Cartesian planes. This non-planarity can signal the presence of matter or a cosmological constant \citep{Podolsky:2003}. Gyratons (spinning relativistic sources) are solutions obtained as a sub-class of Kundt geometries \citep{Frolov:2005,Frolov1:2005,Kadlecova:2009}. Presence of gyratonic matter in a Kundt geometry imparts an angular momentum due to its intrinsic spin. Till date, most of the research around Kundt geometries have largely been focused within the realm of GR \citep{Podolski:2001,Ortaggio:2002,Griffiths:2003,Coley:2009,Podolski:2013,Tahamtan:2017}, though there is some recent work in Gauss-Bonnet \citep{Svarc:2020} and quadratic gravity \citep{Pravda:2017}. As far as we know, not much has been done on such geometries within the ambit of scalar tensor theories. Our article is one such attempt towards (a) contructing new Kundt-type solutions in the simplest scalar-tensor theory, BD gravity and, more importantly,
(b) verifying/contrasting the presence/absence of memory effects 
w.r.t similar ones in GR.

\noindent There does exist previous work on memory effects 
in BD theory which are different from what we wish to 
pursue here. As is well known, in linearised gravity, the BD scalar field produces a breathing mode along with the two additional polarizations ($+,\times$) found in GR  \citep{Will:2014}. Lang computed GW waveforms for scalar and tensor modes separately in the PN approximation \citep{Lang:2014,Lang1:2015}. Du and Nishizawa proposed a test of gravity for scalar tensor theories \citep{Du:2016}. They found two distinct sets of memory contributions: T-memory (tensor) and S-memory (scalar).  Scalar memory is unique in  such theories and was used as a tool to understand the Vainshtein screening mechanism in BD gravity \citep{Koyama:2020}. Asymptotically flat spacetimes in BD theory and memory have been recently studied in \citep{Tahura:2020,Hou:2020}. The BMS group \citep{Bondi:1962} is retained for the tensorial case. There are degenerate vacua for the scalar sector related via Lorentz transformations. The BMS charge algebra is computed in \citep{Hou1:2020}.

\noindent Studying memory effects for such Kundt wave spacetimes in GR was initiated by two of us in \citep{Chak1:2020} through  analysis of geodesic motion. Similar to exact plane wave spacetimes, one can construct sandwich waves here by choosing appropriate limiting profiles \citep{Griffiths:2009,Zhang:2017,Chak:2020}. This serves as a qualitative toy model of a gravitational wave burst. Interesting distinctions occur between negative and positive constant curvature solutions, particularly for the latter, where we found a new {\em frequency memory effect}. In our
analysis here, we consider memory in Kundt geometries without and with gyratonic terms.
First, we construct explicit Kundt solutions for arbitrary $\omega$. The case $\omega = -2$ is 
special because it yields a spacetime with constant negative scalar curvature. For other 
$\omega$  we have variable positive or negative curvature. 
After constructing the solutions, we analyse geodesics with the intent of
studying displacement memory. Thereafter, we move on to geodesic deviation.
We do this by isolating the background, wave and gyratonic contributions to the deviation vector
and their evolution equations. The coupled system of equations are then solved to
obtain the behaviour of the deviation which helps us in analysing the presence of memory.
We will also see how displacement memory is related to the BD scalar field  and 
memory obtained via deviation shows the contributions of the background, gravitational wave 
and/or the gyratonic parts in the total deviation and hence, memory.


  \noindent In our approach towards solving the deviation equation, our calculations are done in Fermi normal coordinates \citep{Chu:2019}. Here the coordinate system is locally Minkowskian and hence the notion of displacement and velocity memory effect is qualitatively similar to exact plane wave spacetimes \citep{Chak:2020,Zhang:2017soft,Shore:2018}. In such Fermi coordinates, we construct parallely transported tetrads along a given timelike geodesic. The deviation vector
  is obtained w.r.t. the tetrad directions and then
  transformed back to the coordinate basis. We demonstrate 
  the calculations in several specific cases (including an
  exactly solvable example) in the relevant section below.
  

\noindent The organization of our paper is as follows. In Sec. II we lay out the basic framework and the tools  used in the paper. Section III deals with Kundt wave solutions without
gyratonic terms and memory effects. Section IV covers the Kundt metric with gyratonic terms 
and aspects of memory. We summarise our work in Sec V with comments on possible extensions.  Relevant 
mathematical formulae used in the paper are listed in an Appendix.
 
\section{Basic framework}

\subsection{Brans-Dicke gravity}

\noindent Brans and Dicke, seeking motivation from Mach's principle, proposed their alternative theory \citep{Brans:1961}, where the Newtonian gravitational constant (G) is related to the reciprocal of a scalar field. This link is based on the idea of variability of inertial mass at different points in spacetime. The action for the BD theory in the Jordan frame is given below. 
\begin{equation}
S=\int \sqrt{-g}\bigg[\phi R -\frac{\omega}{\phi} \nabla_{\alpha}\phi \nabla^{\alpha}\phi +16\pi \mathcal{L}_{m}\bigg] d^4x \label{eq:action}
\end{equation}

\noindent Here, $\phi$ denotes the ambient scalar field, $\omega$ is the BD parameter and the $\mathcal{L}_{m}$ denotes the matter Lagrangian. $\omega$ is a dimensionless parameter which is highly constrained from Solar System observations \citep{Hou:2018}. Different values of $\omega$ correspond to  different theories. 

\noindent  Since its arrival on the scene in the early
sixties, BD theory has been ruled out at times but has reappeared
in different {\em avatars} serving diverse needs. 
For example, its $\omega=-1$
limit is dilaton gravity, which emerges from string theory \citep{Horowitz:1992}.
Similarly, extensions such as replacing a constant $\omega$ with
$\omega(\phi)$ leads to a broader class of theories which
are actively pursued today in cosmological and astrophysical contexts \citep{Faraoni:2004}. Much of the relevance of BD, as well as scalar-tensor
theories, rests on providing templates for comparative studies with 
GR, with the hope of discovering the new physics embodied
in the theoretical constructs or ruling them out. Our work here,
is also an attempt in a similar direction {\em vis-a-vis} the memory effect.

\noindent As mentioned before, we study Kundt geometries \citep{Stephani:2003,Griffiths:2009,Kundt:1961} in BD theory. Given our primary motivation with regard to distinguishing
memory effects arising in GR and BD theory we first construct
the solutions (Kundt-type). In our theoretical setup, we do not restrict the value of $\omega$. Instead we solve the field equations for novel Kundt geometries in BD theory and choose the value of $\omega$ to study specific
cases. Interestingly, we find that in  both the {\em Kundt waves} and {\em gyratonic  Kundt metric}, the value of $\omega$ can be chosen freely. Hence, we analyse memory for specific values $\omega=-2,+1$. The reasons for choosing such specific values are explicitly discussed in section II B 3.  In all the cases considered here, we find vacuum solutions ($\mathcal{L}_{m}=0$).

\noindent The field equations are obtained by variation of $g_{\mu \nu}$ and $\phi$.  After performing a little algebra, we can write them in the standard form as shown below.
\begin{gather}
    G_{\mu \nu}=\frac{\omega}{\phi^2}[\phi,_\mu\phi,_\nu-\frac{1}{2}g_{\mu \nu} \phi,_{\alpha}\phi,^{\alpha}]+\frac{1}{\phi}(\phi,_{\mu;\nu} -g_{\mu\nu}
    \square \phi) \label{eq:metric_var}\\
    \square \phi = 0 \label{eq:phi_var}
\end{gather}
\noindent The box operator is constructed using the Kundt spacetime metric.

\subsection{Kundt geometries and the geodesic analysis of memory}

\noindent Let us first introduce the class known as
as Kundt geometries. We also point out how geodesic analysis leads to our understanding of memory effects for such geometries. Later in Sec. III and IV we provide explicit calculations of memory for various 
solutions in BD theory which are special cases of the
spacetimes mentioned below.

\subsubsection{Kundt wave metric}

\noindent The line element for a Kundt wave geometry is given as,
\begin{equation}
ds^2=-H(u,x,y)du^2-2dudv+ \dfrac{dx^2+dy^2}{P(u,x,y)^2}   \label{eq:metric_Kundt_1} 
\end{equation}

\noindent The waves (denoted via the term $H(u,x,y)$) are viewed as  propagating in
the curved background spacetime \citep{Ortaggio:2002,Podolsky:2003,Chak1:2020}. The background curvature is dependent on $P(u,x,y)$.

\subsubsection{Kundt metric with gyraton/gyraton-like terms}

\noindent The line element for a
generalisation of the metric above is given as, 
\begin{equation}
    ds^2=-Hdu^2-2dudv-2W_1dudx-2W_2dudy+\frac{1}{P^2}(dx^2+dy^2) \label{eq:kundt_metric_3}
\end{equation}
$P\equiv  P(u,x,y), H \equiv H(u,v,x,y), W_i \equiv W_i(u,v,x,y), $ $\forall \hspace{2mm} i  \hspace{2mm}\epsilon \hspace{2mm} \{x,y\}$

\noindent  The vector field $\mathbf{k}=\partial_v$ gives the NGC. The tangent to the spatial surfaces ($P\partial_x,P\partial_y$) and $\mathbf{k}$ are orthogonal to each other. In this paper, we specifically work with gyratonic spacetimes where $W_1,W_2,H$ are independent of coordinate $v$ \citep{Frolov:2005,Kadlecova:2009}. The off-diagonal $W_i$ act as sources of angular momentum in the spacetime and hence, such line elements correspond to spinning null sources. Note that Eq.(\ref{eq:kundt_metric_3}) reduces to  Eq.(\ref{eq:metric_Kundt_1}) if the cross terms ($W_1, W_2$) are set to zero and $H,_v=0$.

\noindent We obtain geometries
in BD theory
representing Kundt waves without and with gyratons 
beginning with metric ansatze given by  Eqs.(\ref{eq:metric_Kundt_1}) or (\ref{eq:kundt_metric_3}) respectively. The functional dependencies of $H$ and $W_i$ 
are chosen as: $H\equiv H(u,x,y)$ and 
$W_i\equiv W_i(u,x,y)$. In both cases, the BD parameter ($\omega$) can be set by hand and is not
constrained by the field equations. Using this freedom we can construct spacetimes with positive or negative curvature 
(Ricci scalar). Thereafter, we study geodesics
and geodesic deviation to infer about memory effects. 

\subsubsection{Geodesic analysis of memory}

\noindent Zhang {\em et al.} \citep{Zhang:2017soft,Zhang:2017} have recently  studied memory effects by analysing the evolution of geodesics in exact plane wave spacetimes. By choosing a Gaussian pulse for the polarization (radiative) term in the line element, they solved the geodesic equations numerically. The change in separation and velocity caused due to the passage of such a pulse was termed displacement and velocity memory effect. We
further extended their analysis on predicting memory from geodesic analysis, for Kundt wave spacetimes in GR \citep{Chak1:2020}.  Expectedly, it was found that there exists a link between memory effects and the
wave profile/curvature of the 
background spacetime. In order to test this hypothesis
in an alternative theory of gravity, we construct and
investigate memory for solutions in BD gravity 
with constant (negative) and varying (positive) curvature scalar. This indicates a direct comparison with results 
on memory in GR within the Kundt class of geometries.

\subsection{Memory effects and geodesic deviation equation}

\noindent Apart from a geodesic analysis, one can understand memory from geodesic deviation. In one of the seminal works on memory effect, Braginsky and Grishchuk \citep{Braginsky:1985} studied geodesic deviation between two test particles moving in a weak gravitational wave. The set of equations following from geodesic deviation was recast into an equation of a forced system. The forcing was shown to be caused due to the passage of the gravitational wave. Integrating the forced equation one could obtain the change in separation between the two particles. This separation was caused only due to the gravitational wave and hence was a residual change ({\em memory}) imprinted on the spacetime.

\noindent In general, the deviation between two neighbouring geodesics in curved spacetimes gets contributions from both gravitational wave part and the non-flat background. Thus, recovering the deviation solely due to the wave (to find the memory effect) is nontrivial. A step along this direction of calculating memory effect in AdS spacetime was initiated in \citep{Chu:2019}. They considered linearized gravitational wave perturbations over AdS  background spacetime. Constructing a parallelly propagated tetrad along a timelike geodesic, they employed Fermi normal coordinates to separate the gravitational wave part from the background. We use the same method as given in \citep{Chu:2019} for our analysis. The only difference here is that in our case the spacetime itself is a radiative geometry while in the former case the perturbations act over a nonradiative background ({\em i.e.} AdS). Fermi coordinates were also used in \citep{Shore:2018} to find the memory for gravitational shock waves and gyratonic pp-waves. Physically it corresponds to a set of timelike inertial observers traversing along the central geodesic whose spatial distance scales are very small compared to the curvature length scale.

\noindent After obtaining the geodesic deviation in the tetrad basis, we transform them back to the coordinate basis. Thus, we determine separately, deviation arising due to different sources ({\em e.g.} background or the wave) in the coordinate basis. The total deviation will, of course, be qualitatively similar to results obtained from the geodesic analysis.


\noindent Let us now explicitly define Fermi normal coordinates and parallelly transported tetrads, which we will use to arrive at memory effects. Consider the geodesic deviation equation
\begin{equation}
    \dfrac{D^2}{d\lambda^2}\xi^{\mu}=-R^{\mu}\,_{\nu\rho\sigma}U^\nu \xi^\rho U^\sigma \label{eq:geodesic_deviation _equation}
\end{equation}

\noindent Here, $\xi^{\mu}$ is the deviation vector between neighbouring timelike geodesics. $U^{\mu}$ is the tangent vector along one of the geodesics. $R^{\mu}\,_{\nu\rho\sigma}$
being the Riemann curvature tensor.  Along a chosen geodesic one can set up a coordinate system $\{t,Z^i\}$ such that the Christoffel connections are always zero along that curve. The spacetime curvature manifests itself through the Riemann curvature. Such basis sets are known as Fermi 
bases \citep{Manasse:1963}. Along the geodesic, $t$ denotes the proper time and $Z^i=0$. A parallely transported tetrad 
is denoted as $e^\mu\,_{\,a}$. This satisfies
\begin{equation}
    U^{\alpha}\nabla_\alpha e^\mu\,_a=0 \label{eq:parallel_tetrad}
\end{equation}

\noindent Construction of such a tetrad 
depends on the spacetime metric. In the tetrad $e^\mu\,_a$, the Greek indices are the spacetime coordinates while Latin indices are for the Fermi coordinates. We have
$e^\mu\,_0=U^\mu$ (tangent vector)  due to the geodesic equation. In the vicinity of the central geodesic, a neighbouring geodesic having separation $\xi^\mu$ 
is re-written in the Fermi basis as,
\begin{equation}
    \xi^\mu=Z^i e^\mu\,_i \label{eq:dev_tetrad}
\end{equation}

\noindent Hence, the geodesic deviation equation in  Fermi coordinates $(t=X^0,X^i)$ becomes:
\begin{equation}
    \frac{d^2Z^i}{dt^2}=-{R}^i\,_{0j0} Z^j \label{eq:deviation_eqn1}
\end{equation}

\noindent $t$ and $\lambda$ both being affine parameters are related via affine transformations. Eq.(\ref{eq:deviation_eqn1}) can be obtained by substituting Eq.(\ref{eq:dev_tetrad}) (and also using  Eq.(\ref{eq:parallel_tetrad})) in  Eq.(\ref{eq:geodesic_deviation _equation}). The spatial indices associated with the frame are denoted by $i,j$ (${R}^i\,_{0j0}={R}^\mu\,_{\nu\rho\sigma}e^i\,_\mu e^\nu\,_0 e^\rho\,_j e^\sigma\,_0$). The tetrads and metric are related via \footnote{$\eta_{ij}$ denotes the Minkowski metric with signature (-1,1,1,1).}$\eta_{ij}=e^\alpha\,_ie^\beta\,_jg_{\alpha\beta}$.

 \noindent Since the deviation has contributions both from background and gravitational radiation, we assume that the total deviation vector is decomposed in the form: $Z^i=Z^i_B+Z^i_W$, where the suffixes $B,W$ are for background and wave respectively. As already mentioned, a similar analysis has been carried out in \citep{Chu:2019} to separate the radiation from background curvature. The splitting of the Riemann tensor is done by noting the terms which are proportional to $H(u,x,y)$ or its derivatives ( ${R}^i\,_{0j0}=({R}^i\,_{0j0})_B+({R}^i\,_{0j0})_W$). Such terms denote the gravitational wave contribution while the other terms are due to background curvature. Thus, Eq.(\ref{eq:deviation_eqn1}) separates into the two equations shown below.
\begin{gather}
    \frac{d^2Z^i_B}{dt^2}=-({R}^i\,_{0j0})_B Z^j_B \label{eq:deviation_eqn1_bg}\\
    \frac{d^2Z^i_W}{dt^2}=-[({R}^i\,_{0j0})_B+({R}^i\,_{0j0})_W] Z^j_W-({R}^i\,_{0j0})_W Z^j_B\label{eq:deviation_eqn1_wave}
\end{gather}
\noindent  Eq.(\ref{eq:deviation_eqn1_bg}) is the deviation due to the background. This would have been the complete geodesic deviation equation if $H(u,x,y)=0$ in the metric line element as given in Eq.(\ref{eq:metric_Kundt_1}). Solving  Eq.(\ref{eq:deviation_eqn1_wave}) gives the memory effect in the tetrad frame. Once both background and wave deviation ({\em i.e.}$Z^i_B$ and $Z^i_W$) are known, we revert back to the coordinate basis using Eq.(\ref{eq:dev_tetrad}) to obtain $\xi_B^{\mu}$, $\xi_W^{\mu}$ and  $\xi^{\mu}(=\xi_B^{\mu}+\xi_W^{\mu})$.

\noindent  In the case of gyratons, we split the deviation vector as: $Z^i=Z^i_B+Z^i_G+Z^i_W$. Apart from the familiar terms $Z^i_B$ and $Z^i_W$, we also have deviation due to gyratons denoted by $Z^i_G$. The Riemann tensor corresponding to this deviation comprises of terms linked to $W_i(u,x,y)$ given in Eq.(\ref{eq:kundt_metric_3}). The deviation equations corresponding to the three distinct effects are given by the Eqs.(\ref{eq:deviation_bg_gyraton}), (\ref{eq:deviation_gyraton_gyraton}) and (\ref{eq:deviation_wave_gyraton}) in Sec. IV B. We then go on to calculate $\xi_B^{\mu},  \xi_G^{\mu}$ and $\xi_W^{\mu}$ in coordinate basis. The terms $\xi_G^{\mu}$ and $\xi_W^{\mu}$ correspond to coordinate memory effect for the gyraton and the wave respectively.

\noindent We have already emphasized the advantage of using Fermi-normal coordinates for our analysis. An important feature is that Eqs.(\ref{eq:deviation_eqn1}), (\ref{eq:deviation_eqn1_bg}) and (\ref{eq:deviation_eqn1_wave}) take their respective forms only when the constructed tetrads are parallely transported.

\noindent It is also important to note that the results on memory from geodesic deviation
are expected to match only qualitatively with those obtained from a geodesic analysis.
This is because the deviation equation is perturbative by construction. In contrast there
is no such restriction when we consider pairs of geodesics and differences in their separation caused by a pulse.


\section{Kundt wave metric}

\noindent We now focus on finding solutions 
in BD theory which represent Kundt wave spacetimes 
generically given as in  Eq.(\ref{eq:metric_Kundt_1}). 
 The BD scalar field is assumed to be independent of $v$ and hence, $\phi\equiv \phi(u,x,y)$. The components  of Eq.(\ref{eq:metric_var}) which are relevant for solving the field equations are listed below.\footnote{The four other equations are the redundancies of the Einsteins field equations and hence are not required for obtaining the solutions.}
\begin{gather}
    G_{xx}=\frac{\omega}{2\phi^2}(\phi,_x^2-\phi,_y^2)+\frac{1}{\phi}(\phi,_{xx}+\frac{P,_x}{P}\phi,_x-\frac{P,_y}{P}\phi,_y) \label{eq:xx}\\
    G_{yy}=\frac{\omega}{2\phi^2}(\phi,_y^2-\phi,_x^2)+\frac{1}{\phi}(\phi,_{yy}+\frac{P,_y}{P}\phi,_y-\frac{P,_x}{P}\phi,_x) \label{eq:yy}\\
    G_{uu}=\frac{\omega}{\phi^2}\bigg(\phi,_u^2+\frac{H}{2}P^2(\phi,_x^2+\phi,_y^2)\bigg)+\frac{1}{\phi}\bigg(\phi,_{uu}-\frac{1}{2}P^2H,_x\phi,_x-\frac{1}{2}P^2H,_y\phi,_y\bigg) \label{eq:uu}\\
    G_{uv}=\frac{\omega}{2\phi^2}P^2(\phi,_x^2+\phi,_y^2)\label{eq:uv}\\
    G_{xu}=\frac{\omega}{\phi^2}(\phi,_x\phi,_u)+\frac{1}{\phi}\bigg(\phi,_{xu}+\frac{P,_u}{P}\phi,_x\bigg) \label{eq:xu}\\
    G_{yu}=\frac{\omega}{\phi^2}(\phi,_y\phi,_u)+\frac{1}{\phi}\bigg(\phi,_{yu}+\frac{P,_u}{P}\phi,_y\bigg) \label{eq:yu}
\end{gather}

\noindent We decompose the scalar field and the metric functions as:
\begin{equation}
    \phi(u,x,y)=\alpha(u)\psi(x,y), \hspace{0.5cm} P(u,x,y)=\dfrac{\tilde P(x,y)}{U(u)}, \hspace{0.5cm}
    H(u,x,y)=H'(u)h(x,y). \label{eq:decom_2}
\end{equation}

\noindent Adding equations (\ref{eq:xx}) and (\ref{eq:yy}) and using separation of variables from equation (\ref{eq:decom_2}) results in $\psi,_{xx}+\psi,_{yy}=0$ ( we know that $G_{xx}=G_{yy}=0$ from the metric). The solution is,
\begin{equation}
    \psi(x,y)=a+\log(x^2+y^2) \label{eq:psi}
\end{equation}
\noindent From the metric, $G_{uv}=P^2\Delta\log P$ (where, $\Delta=(\partial_{xx}+\partial_{yy})$).  Using this in equation (\ref{eq:uv}) gives,
\begin{equation}
    \tilde P= \frac{\sqrt{x^2+y^2}}{[a+\log(x^2+y^2)]^{\omega/2}} \label{eq:P_kw}
\end{equation}

\noindent Eq.(\ref{eq:P_kw}) shows that background curvature is explicitly dependent on $\omega$. The equations for the `{\em xu}' and `{\em yu}' components, as in (\ref{eq:xu}),(\ref{eq:yu}) and given the metric (\ref{eq:metric_Kundt_1}) we end up with
\begin{align}
    \frac{\omega}{\phi^2}\big(\phi,_x\phi,_u)+\frac{1}{\phi}(\phi,_{xu}+\frac{P,_u}{P}\phi,_x\big)=\bigg(\frac{P,_u}{P}\bigg),_x \label{eq:xu_eqn}\\
    \frac{\omega}{\phi^2}\big(\phi,_y\phi,_u)+\frac{1}{\phi}(\phi,_{yu}+\frac{P,_u}{P}\phi,_y\big)=\bigg(\frac{P,_u}{P}\bigg),_y \label{eq:yu_eqn}
\end{align}

\noindent Both the above equations reduce to the same equation after using the separation of variables. We have,
\begin{equation}
    (\omega+1)\frac{\alpha,_u}{\alpha}=\frac{U,_u}{U} \label{eq:relation}
\end{equation}

\noindent The Ricci scalar curvature is
\begin{equation}
    R=2P^2\Delta \log P=\frac{4\omega}{U^2[a+\log(x^2+y^2)]^{\omega+2}} \label{eq:scalar}
\end{equation}

\noindent The component of $G_{uu}$ from the metric is given below. 
$$G_{uu}=\frac{P^2}{2}(H,_{xx}+H,_{yy})+2\frac{P,_{uu}}{P}-4\bigg(\frac{P,_u}{P}\bigg)^2+H(-P,_x^2-P,_y^2+P(P,_{xx}+P,_{yy}))$$

\noindent Using equations (\ref{eq:uu}) and (\ref{eq:decom_2}) we get 
\begin{equation} \label{eq:uu_eqn}
\begin{split}
   & \frac{\tilde P^2}{2U^2}H'(u)(h,_{xx}+h,_{yy})-2\frac{U,_{uu}}{U}+H'(u)h\bigg(\frac{\tilde P(\tilde P,_{xx}+\tilde P,_{yy})-\tilde P,_x^2- \tilde P,_y^2}{U^2}\bigg)=
    \\
    & \omega \bigg[\bigg(\frac{\alpha_{ ,u}}{\alpha}\bigg)^2+\frac{H'h\tilde P^2}{2U^2}\bigg(\bigg(\frac{\psi,_x}{\psi}\bigg)^2+\bigg(\frac{\psi,_y}{\psi}\bigg)^2\bigg)\bigg]+\frac{\alpha,_{uu}}{\alpha}-\frac{\tilde P^2}{2U^2}H'(u)\bigg(h,_x\frac{\psi,_x}{\psi}+h,_y\frac{\psi,_y}{\psi}\bigg)
    \end{split}
\end{equation}

\noindent We set $U=1$. Hence, from Eq.(\ref{eq:relation}) we get $\alpha(u)$ as a constant . From equations (\ref{eq:relation}), (\ref{eq:uu_eqn}) and  (\ref{eq:psi}) we find that $H'(u)$ is unconstrained. The $xy$ dependent part of $H(u,x,y)$ becomes
\begin{equation}
    h(x,y)=\log[a+\log(x^2+y^2)] \label{eq:h}
\end{equation}

\noindent This polarization term $h(x,y)$ is different from GR. We will point out the consequences of this difference on the nature of the memory effect, contrasting it with GR. Thus from our generic analysis  we find that only $H'(u)$ and $\omega$ is unconstrained. 

\noindent We now perform a coordinate transformation $x=e^{X-a/2}\cos Y, y=e^{X-a/2} \sin Y$. The metric in the new coordinates $(u,v,X,Y)$ becomes
\begin{equation}
    ds^2= -H'(u)\log (2X)du^2-2dudv +(2X)^\omega (dX^2+dY^2) \label{eq:Kundtwave_gen_omega_metric}
\end{equation}

\noindent The reverse transformation shows that $X=\frac{1}{2}(a+\log[(x^2+y^2)])\equiv\frac{1}{2}\psi[X]$. This relation shows the imprint of the scalar field in the solution of the metric, via coordinate $X$. Thus, $\phi[X(u)]$ evaluated along the geodesic $X(u)$ gives a measure of the memory effect for the scalar field due to the gravitational wave pulse.

\subsection{Displacement memory using geodesics}

\noindent {The geodesic equations of coordinates $X,Y$ for the metric line element in Eq.(\ref{eq:Kundtwave_gen_omega_metric}) are given below.
\begin{gather}
    \dfrac{d^2X}{du^2}+\frac{\omega}{2X}\bigg[\bigg(\dfrac{dX}{du}\bigg)^2-\bigg(\dfrac{dY}{du}\bigg)^2\bigg]+\frac{H'(u)}{(2X)^{\omega+1}}=0 \label{eq:X_geod_gen_omega}\\
    \dfrac{d^2Y}{du^2}+\frac{\omega}{X}\bigg(\dfrac{dX}{du}\bigg)\bigg(\dfrac{dY}{du}\bigg)=0 \label{eq:Y_geod_gen_omega}
\end{gather}

\noindent Geodesic equations for coordinate $v$ is trivial ($\ddot{u}=0$). Hence, $u$ acts as an affine parameter. We try to solve  Eqs.(\ref{eq:X_geod_gen_omega}) and (\ref{eq:Y_geod_gen_omega}) by setting the initial value of the transverse coordinate velocities to zero (\textit{i.e.} $\dot{X}=\dot{Y}=0$). Taking $\dot{Y}=0$ in Eq.(\ref{eq:Y_geod_gen_omega}) gives $\ddot{Y}=0$. Hence, $\dot{Y}=0$ for the entire evolution of the geodesic. The only non-trivial equation left 
is for the coordinate $X$ which is given as,
\begin{equation}
    \frac{d^2X}{du^2}+\frac{\omega}{2X}\bigg(\dfrac{dX}{du}\bigg)^2+\frac{H'(u)}{(2X)^{\omega+1}}=0 \label{eq:gen_X_geo_kwave}
\end{equation}

\noindent We use the transformation $X=\frac{1}{2}q^{\frac{1}{\omega+2}}$
for $\omega\neq 2$. Eq.(\ref{eq:gen_X_geo_kwave}) is thus transformed to the equation
\begin{equation}
    \frac{d^2q}{du^2}-\frac{1}{2q}\bigg(\frac{dq}{du}\bigg)^2+2(\omega+2)H'(u)=0 \label{eq:levinson_smith}
\end{equation}
\noindent The above equation resembles that of a forced Levinson-Smith system which, generically, has an equation of the form \cite{Levinson:1942} \citep{Krishchenko:2019},
\begin{equation}
\ddot x + a(x,\dot x) \dot x + g(x) = f(t)
\end{equation}
where $a(x,\dot x)$, $g(x)$ and $f(t)$ need to be specified.
A comparison with the equation for $q(u)$ given just above,
shows the correspondence. Explicit solutions  are not quite available,
especially for the equation in our case, though a dynamical systems
analysis exists \citep{Krishchenko:2019}.

\noindent The nature of the forcing term $H'(u)$ (which
encodes the effect of the gravitational wave pulse) dictates the
behaviour of any solution. 
Therefore, it is likely that the analysis of such forced equations with nonlinearities 
may play a role in understanding the memory effect. Related discussion
on the relevance of a forced equation appeared much earlier (in the context of the deviation
equation) in the original 
work on memory by Braginsky-Grishchuk \citep{Braginsky:1985}. 
Thus, this brief side remark on the similarity of the 
geodesic equation for $X$ (for $\omega\neq -2$) with a known, nonlinear forced system, as noted 
above, seems worth investigating further, in future. 
}

\noindent {In principle, the geodesic equation given in Eq.(\ref{eq:gen_X_geo_kwave}) can be solved numerically for any value of $\omega$ for a given choice of the pulse $H'(u)$. We have chosen $H'(u)=\frac{1}{2}\sech^2 u$ 
since it qualitatively resembles a gravitational wave pulse. However, as mentioned earlier, we are interested in observing memory effects for different choices of $\omega$ corresponding to different scalar curvature scenarios.}
 The value $\omega=-2$ is special since it is the only case where
the Ricci scalar is constant and negative. For other $\omega$ one may have variable positive or negative $R$. We have chosen 
to illlustrate our analysis for the $\omega \neq -2$ case 
with a $\omega$ value which yields
a positive but variable Ricci scalar. Results for variable negative Ricci scalar are not very different and can be worked out easily too. We discuss each case briefly with corresponding plots.

\begin{itemize}
    \item {\underline{$\omega=-2$}\\
\noindent {We consider a scenario where the scalar curvature is constant. Hence, $\omega=-2$ and the scalar
curvature is $R=-8$. The geodesic equation (\ref{eq:gen_X_geo_kwave}) becomes}
\begin{equation}
    \frac{\ddot{X}}{X}-\bigg(\frac{\dot{X}}{X}\bigg)^2+\sech^2(u)=0 \label{eq:X_geodesic_-2_kw}
\end{equation}
One can solve Eq.(\ref{eq:X_geodesic_-2_kw}) analytically. Setting $\dot{X}/{X}=p$, one finds that
\begin{equation}
    p(u)=-(1+\tanh(u)) \label{eq:reflection}
\end{equation}
The constant is fixed by setting $p=0$ at $u \to -\infty$ as initially $\dot{X}$ also vanishes. 
Solving $X$ from the analytical form of $p(u)$ yields
\begin{equation}
    X(u)=\frac{A}{1+e^{2u}} \label{X_soln_-2_kw}
\end{equation}
{Here, $A$ denotes the initial position of the particle. One can check that as $u\to -\infty$, $X=A$ whereas as $u\to+\infty$, $X=0$. Thus, two geodesics  starting with different initial coordinate values, eventually have a zero $X$ value, after the passage of the pulse.}

\noindent {Let us now consider two different geodesics having initial X coordinate value as $A_1$ and $A_2$.   We find, as $u\to-\infty$, $X_1=A_1$, $X_2=A_2$. The change in initial separation is $X_1-X_2=A_1-A_2$. The final separation at $u\to+\infty$ is zero . Thus. we have displacement
memory for the coordinate $X$. }

\noindent {A careful inspection reveals that Eq.(33) 
is invariant under $u$ $\rightarrow -u$.
Hence, the
analytical solution,  
$X(u)=\frac{A}{1+e^{-2u}}$
is also possible. 
Here,
two geodesics 
both
starting from $X=0$ settle to
two different final values (depending on $A_1-A_2$). We illustrate 
both
the analytical results below using plots.}

 \begin{figure}[H]
	\centering
	\begin{subfigure}[t]{0.32\textwidth}
		\centering
		\includegraphics[width=\textwidth]{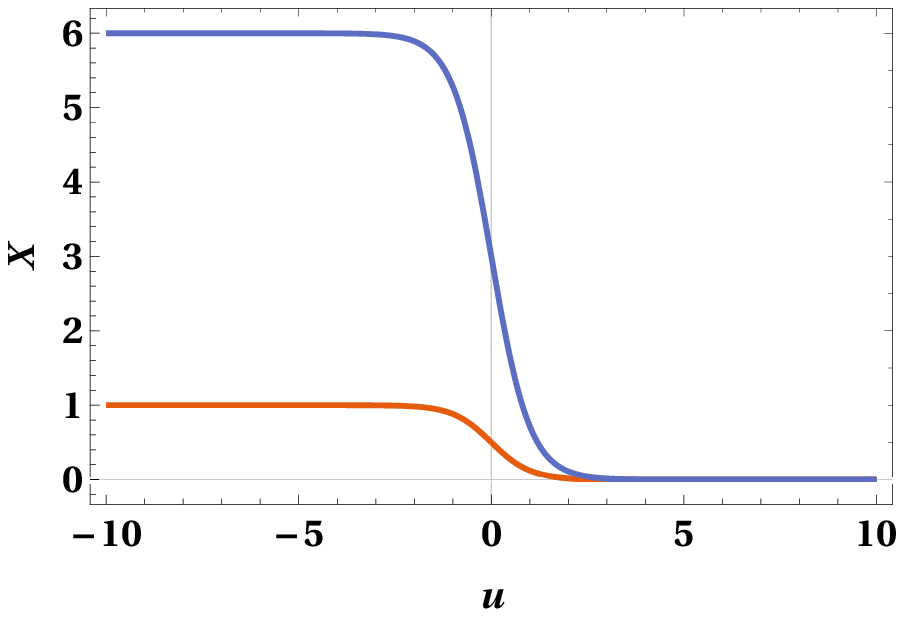}
		\caption{ \centering Geodesics with initial position as $A_1=1,A_2=6$.}
		\label{fig:Kundt_X_-2}
\end{subfigure}\hspace{1.5cm}
	\begin{subfigure}[t]{0.32\textwidth}
		\centering
		\includegraphics[width=\textwidth]{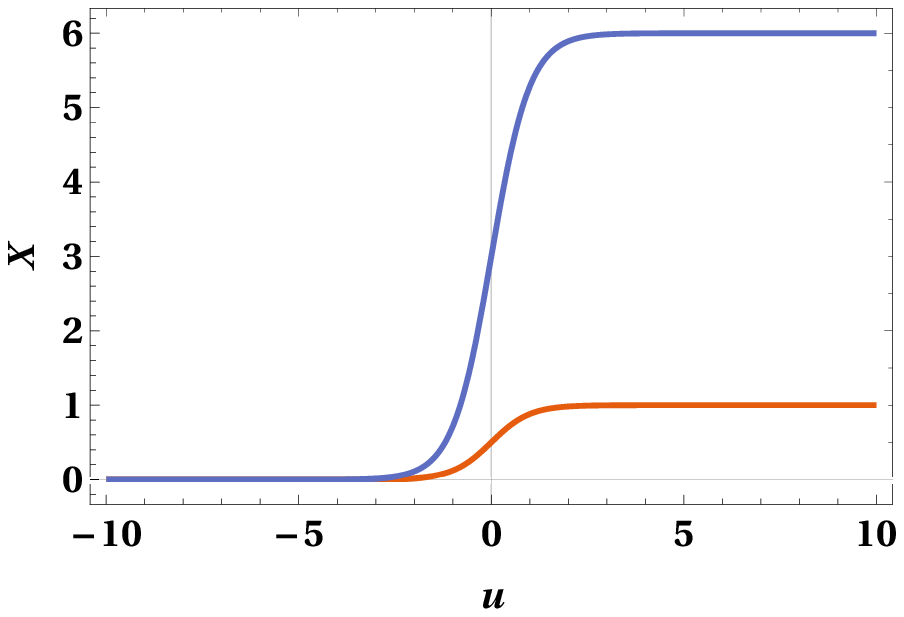}
		\caption{\centering Geodesics with final position as $A_1=1,A_2=6$.}
		\label{fig:reflection_-2}
	\end{subfigure}
	\caption{Displacement memory effect for Kundt waves with $\omega=-2$.}
	\label{fig:Kundtwave_-2}
\end{figure}

\noindent {We observe permanent displacement along X-direction (see Fig.(\ref{fig:Kundtwave_-2})). This is qualitatively similar to GR ( 
see
\citep{Chak1:2020}) where we also observed constant separation after the passage of the gravitational wave pulse.  Thus, for negative scalar curvature solutions, there is a
qualitative agreement in the nature of the memory effect
as found in GR and in BD ($\omega=-2$) 
theory.
}
\item {\underline{$\omega\neq-2$}\\
The earlier geodesic analysis reveals that constant negative curvature case of $\omega=-2$ is integrable. 
We have not been able to find analytical solutions for other values of $\omega$. Resorting to numerics, we study the
behaviour of the coordinate $X$ for $\omega=+1$. The motivation for choosing $\omega=+1$ is to compare between the results on memory for positive curvature solutions as obtained here in BD theory with those in GR. The Ricci scalar 
for $\omega=+1$ becomes $4/(8X^3)$. Thus, there is a possibility of having negative scalar curvature in $\omega=+1$ by choosing coordinate ranges where $X<0$. However, in our entire analysis of memory effects, we have restricted our coordinate range beyond the singular region so that the solution conforms to the positive sign of scalar curvature. 
We have thus avoided the negative scalar curvature region. Nevertheless, we have observed that the singularity does 
seem to influence pairs of geodesics through their gradual
convergence towards it.} 

Eq.(\ref{eq:gen_X_geo_kwave}) is solved numerically in { \em Mathematica 10} and the evolution of coordinate $X$ is shown below.

\begin{figure}[H]
	\centering
		\includegraphics[scale=0.7]{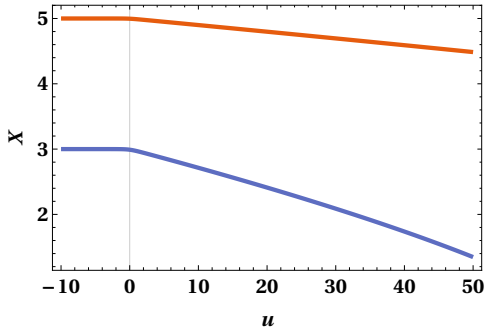}
		\caption{\small $\omega=1$: Initial position of $X$ for the two geodesics are 5(orange)and 3(blue) respectively.}
		\label{fig:x_1_geodesic}
\end{figure}

In Fig.(\ref{fig:x_1_geodesic}), we find increasing separation between the geodesics after the departure of the pulse. This is in sharp contrast to the profiles obtained in GR. In the latter theory we found from geodesic analysis that positive curvature scenarios give rise to a {\em frequency memory effect} \citep{Chak1:2020}. This is related to the different  metric functions in the Kundt wave line element for the two theories. In BD theory, $h(X,Y)=\log(2X)$ (obtained by solving the field equations) whereas in GR we took it as $h(X,Y)=\frac{1}{2}(X^2-Y^2)$ (usual expression found in $+$ polarization). We also find that the geodesics do not cross beyond $X=0$. This is due to the singular nature of the metric as mentioned just above. 
}
\end{itemize}

\subsection{Geodesic deviation analysis of memory}

\noindent We now turn towards discussing geodesic deviation
following the method outlined in Section II C. First, we  construct an orthonormal tetrad for the metric line element (\ref{eq:Kundtwave_gen_omega_metric}).   
\begin{equation}
\begin{split}
 e_0\,^{\mu}=[1,\dot{v},\dot{X},\dot{Y}] \hspace{1.5cm} 
    e_1\,^{\mu}=[0,-(2X)^{\omega/2}\dot{X},-(2X)^{-\omega/2},0] \\
 e_2\,^{\mu}=[0,-(2X)^{\omega/2}\dot{Y},0,-(2X)^{-\omega/2}]  \hspace{1.5cm} 
    e_3\,^{\mu}=[-1,1-\dot{v},-\dot{X},-\dot{Y}] \label{eq:tetrad_withoutc}
    \end{split}
\end{equation}

\noindent A similar construction was 
carried out in \citep{Bicak:1999} for a different coordinate system. $e_0\,^{\mu}$ gives the tangent to the geodesic. The parallel transport condition (\ref{eq:parallel_tetrad}) is only satisfied by $e_0\,^{\mu}$ (obeys the geodesic equations) and $e_3\,^{\mu}$. Both $e_1\,^{\mu},e_2\,^{\mu}$  are not parallely transported. Hence, these two tetrads are rotated by an angle $\dot{\theta}_p=\omega\dot{Y}/(2X)$. Since we have $\dot{Y}=0$ for all $\omega$ from the geodesic analysis, $\theta_p$ is a constant.  We take $\theta_p=0$ so that the two tetrads $e_1\,^{\mu},e_2\,^{\mu}$ also satisfy Eq.(\ref{eq:parallel_tetrad}).

\noindent The non-zero Riemann tensor components in the tetrad basis are shown below.

\noindent {\em Background}
\begin{equation}
\begin{split}
(R^1\,_{010})_B=\frac{\omega\dot{Y}^2}{2X^2} \hspace{2.5cm}
(R^1\,_{020})_B=-\frac{\omega\dot{Y}\dot{X}}{2X^2}\\ 
(R^2\,_{010})_B=-\frac{\omega\dot{Y}\dot{X}}{2X^2}  \hspace{2.5cm}
(R^2\,_{020})_B=\frac{\omega\dot{X}^2}{2X^2}
 \label{eq:riemann_tetrad_gen_omega_bg}
    \end{split}
\end{equation}

\noindent  {\em Wave}
\begin{equation}
(R^1\,_{010})_W=-\frac{(\omega+2) H'(u)}{(2X)^{\omega+2}} \hspace{2.5cm}
(R^2\,_{020})_W=\frac{\omega H'(u)}{(2X)^{\omega+2}}
 \label{eq:riemann_tetrad_gen_omega_wave}
\end{equation}

\noindent Substituting the expressions in Eq.(\ref{eq:riemann_tetrad_gen_omega_bg})  and Eq.(\ref{eq:riemann_tetrad_gen_omega_wave}) (for the Riemann tensor components in the tetrad basis) in Eqs.(\ref{eq:deviation_eqn1_bg}) and (\ref{eq:deviation_eqn1_wave}), we solve for the background and gravitational wave contributions to the geodesic deviation.

\noindent As pointed out earlier, we ultimately go over to the coordinate basis using Eq.(\ref{eq:dev_tetrad}). We find out $\xi_B^\mu, \xi_W^\mu$ only along $X,Y$ directions. The total deviation $\xi^\mu$ is then obtained and compared with the results obtained from geodesics.

\noindent This decomposition of deviation vectors into background and wave parts can be done because the geodesic deviation equation is {\em linear}. As the geodesic equations itself are {\em non-linear}, this method of decomposition is not possible. One may also directly integrate the geodesic deviation equation in the coordinate basis and come to similar conclusions as ours. However, using the tetrads, the equations simplify enough, as can be seen by comparing Eqs.(\ref{eq:geodesic_deviation _equation}) and (\ref{eq:deviation_eqn1}). Interestingly, for $\omega=-2$ we have an exact solution. 

\noindent We now show that by
solving the geodesic deviation equation we may obtain the memory effect.
 The change in the part of the deviation vector related to the background arises due to the $P(u,x,y)$ term in Eq.(\ref{eq:metric_Kundt_1}). The change caused by the pulse $H(u,x,y)$ is
conventionally related to memory, largely because the
pulse is viewed as the `cause'. Our choice of Fermi coordinates 
simplifies the calculations to some extent, though it is 
surely possible to do everything in the coordinate basis as
well.

\noindent  Below, we use the previously chosen $\omega$ values
(as in the geodesic analysis) to carry out the deviation 
analysis. Since $u$ is an affine parameter we can replace the proper time $t$ mentioned earlier, with $u$.

 \noindent \underline{$\omega=-2$:}

\vspace{0.1in}

 \noindent As we have seen, the geodesic equations are analytically solvable in this case. We will use the solution $X=\dfrac{A}{1+e^{-2u}}, \dot{Y}=0$ for further investigation. In this scenario, two geodesics starting from zero initial value have two different final separations (see Fig.(\ref{fig:reflection_-2})). Substituting $\omega=-2$ in  Eqs.(\ref{eq:riemann_tetrad_gen_omega_bg}) and (\ref{eq:riemann_tetrad_gen_omega_wave}) gives a nontrivial equation only along $Z^2$ direction. We have,
\begin{gather}
      \ddot{B}= [1-\tanh(u)]^2 B \label{eq:B_omega_-2}\\
      \ddot{W}=2[1-\tanh(u)] W+ \sech^2(u)B \label{eq:W_omega_-2}\\
      \ddot{K}=2[1-\tanh(u)]K \label{eq:K_omega_-2}
\end{gather}
 \noindent where $B=Z_B^2, W=Z_W^2, K=B+W$. The analytical solutions for Eqs.(\ref{eq:B_omega_-2}) and (\ref{eq:K_omega_-2}) are
 \begin{equation}
 \begin{split}
     B(u)=C_1(1+e^{-2u})^{\frac{1-\sqrt{5}}{2}}  {}_2F_1\bigg[-\frac{1+\sqrt{5}}{2}, \frac{3-\sqrt{5}}{2}, 1 - \sqrt{5}, -(1 + e^{-2 u})\bigg] \\
     +C_2(1+e^{-2u})^{\frac{1+\sqrt{5}}{2}}  {}_2F_1\bigg[\frac{\sqrt{5}-1}{2}, \frac{3+\sqrt{5}}{2}, 1+\sqrt{5}, -(1 + e^{-2 u})\bigg]
     \label{eq:B_soln}
 \end{split}
 \end{equation}

\begin{equation}
     K(u)=\frac{1+e^{-2u}}{C_3} \label{eq:K_soln}
\end{equation} 
 
 \noindent The solutions for $B(u)$ has hypergeometric functions denoted by ${}_2F_1$. $C_1,C_2,C_3$ are the constants of integration. We set $\dot{K}=0$ as $u\to -\infty$ to get rid of the other constant. $W(u)$ can be easily obtained by subtracting $B(u)$ from $K(u)$. Since we start with zero initial velocities, we find $Z^1_B$ and $Z^1_W$ are constants. Eq.(\ref{eq:K_soln}) shows that the total deviation $K(u)$ is inverse of the geodesic solution $X(u)$.

\noindent Reverting back to the coordinate basis, the deviation along coordinates $X$ and $Y$ becomes
\begin{gather}
    \xi^X_B= (-2X)Z^1_B \hspace{3.5cm} \xi^X_W=(-2X)Z^1_W \label{eq:xi_X}\\
    \xi^Y_B=(-2X)B \hspace{3.8cm} \xi^Y_W=(-2X)W \label{eq:xi_Y}
\end{gather}

\begin{figure}[H]
 \centering
	\begin{subfigure}[t]{0.45\textwidth}
		\centering 
		\includegraphics[width=\textwidth]{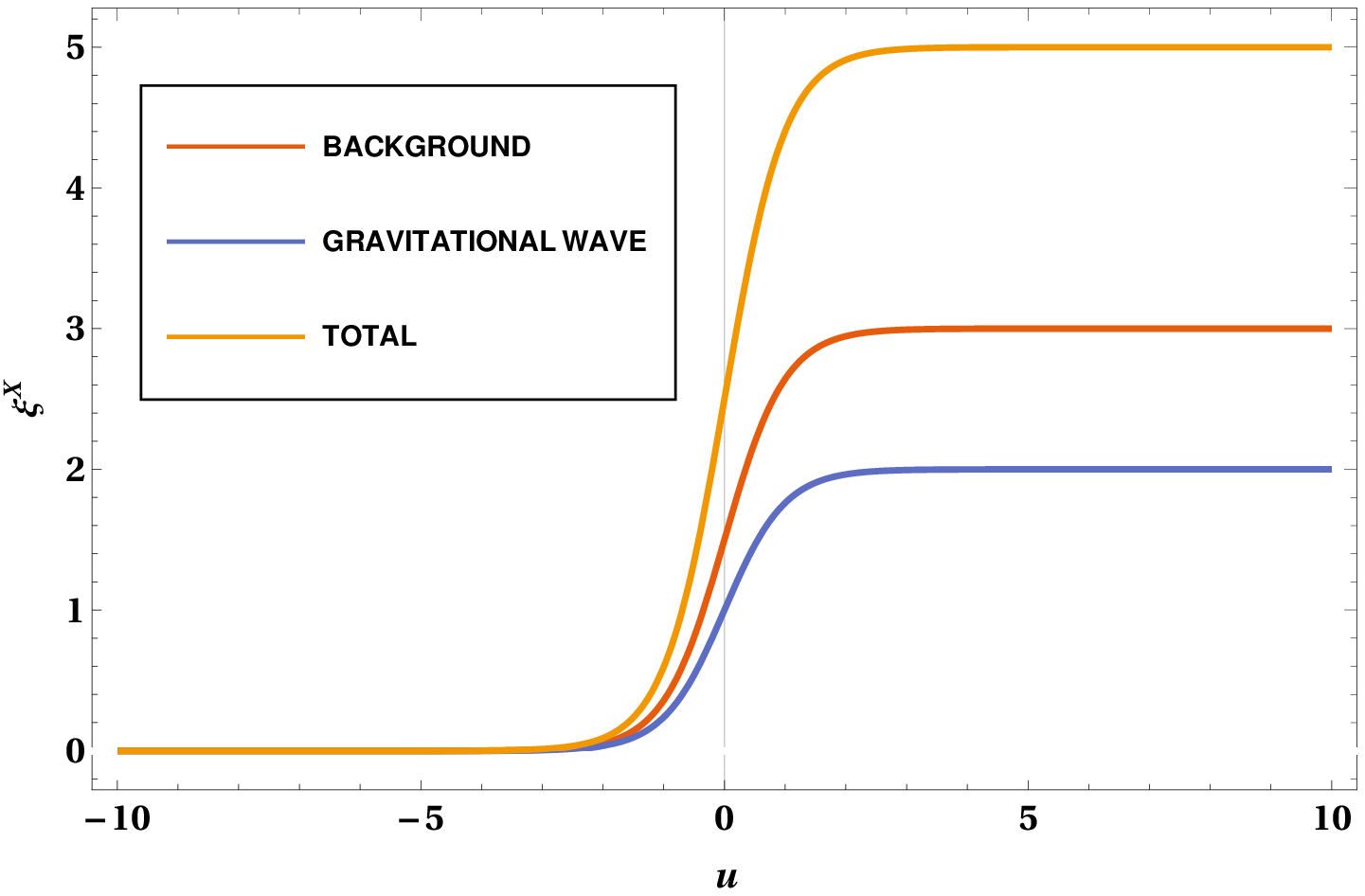}
		\caption{$\xi^X$ vs $u$}
		\label{fig:xi_X_omega_-2}
\end{subfigure}\hspace{1cm}
	\begin{subfigure}[t]{0.45\textwidth}
		\centering
		\includegraphics[width=\textwidth]{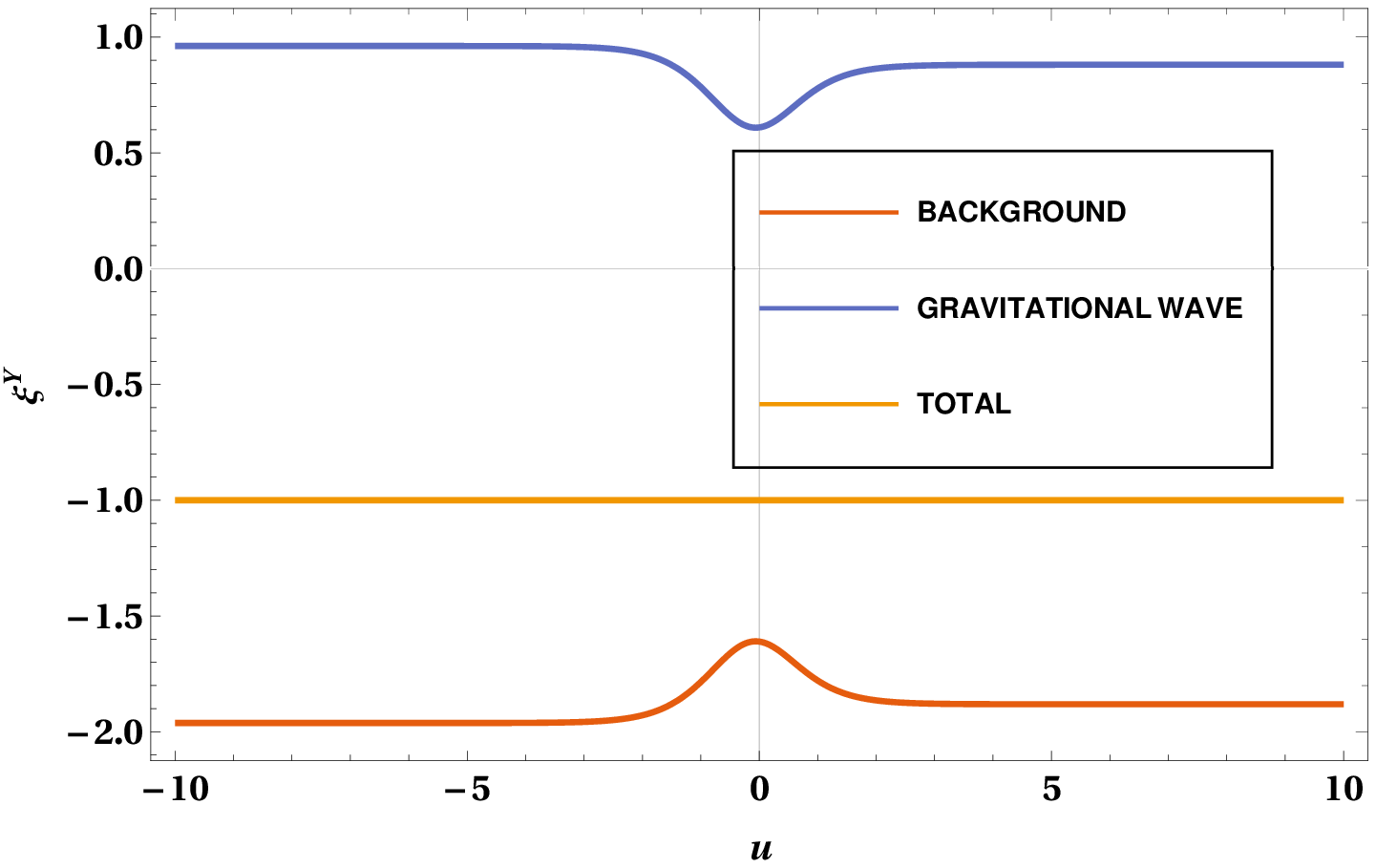}
		\caption{$\xi^Y$ vs $u$}
		\label{fig:xi_Y_omega_-2}
	\end{subfigure}
	\caption{Plot of deviation vectors along the coordinates $X$ and $Y$ for $\omega=-2$.}
	\label{fig:deviation_-2_coo}
\end{figure}

\noindent Eq.(\ref{eq:xi_X}) shows how both the background and the wave deviation are simply proportional to $X$. This also follows from the Fig.(\ref{fig:xi_X_omega_-2}). Thus, the nature of the three deviations (background, wave and total) are all similar. The background and the gravitational wave both sum up to enhance the amplitude of the total deviation. Along the $Y$ direction in Fig.(\ref{fig:xi_Y_omega_-2}), we observe that the wave and background deviations cancel each other. The total deviation is a constant. Thus, the geodesic deviation analysis also gives the same qualitative result on separation, as obtained from the geodesics. This confirms our previous assertion that the geodesic analysis can only 
`see' a total deviation (or separation). It is unable to retrieve the gravitational wave contribution from the total. In both plots (Fig.(\ref{fig:deviation_-2_coo})), the {\em blue} line shows the gravitational wave memory effect.

\vspace{0.1in}
\noindent \underline{$\omega\neq-2$:}
\vspace{0.1in}

\noindent Similar to the geodesic analysis, we perform the deviation analysis for $\omega=+1$, which corresponds to a spacetime with a positive (but varying)
Ricci scalar. The deviation equation is solved numerically. We follow the same steps as we did for $\omega=-2$. From Eq.(\ref{eq:riemann_tetrad_gen_omega_bg}), we get that $Z_B^1$ is a constant. The evolution equations of $Z_B^2$, $Z_W^1$ and $Z_W^2$ are solved numerically in {\em Mathematica 10}. 
Finally, we go over to the coordinate basis to state our results on memory.

 \begin{figure}[H]
 \centering
	\begin{subfigure}[t]{0.45\textwidth}
		\centering 
		\includegraphics[width=\textwidth]{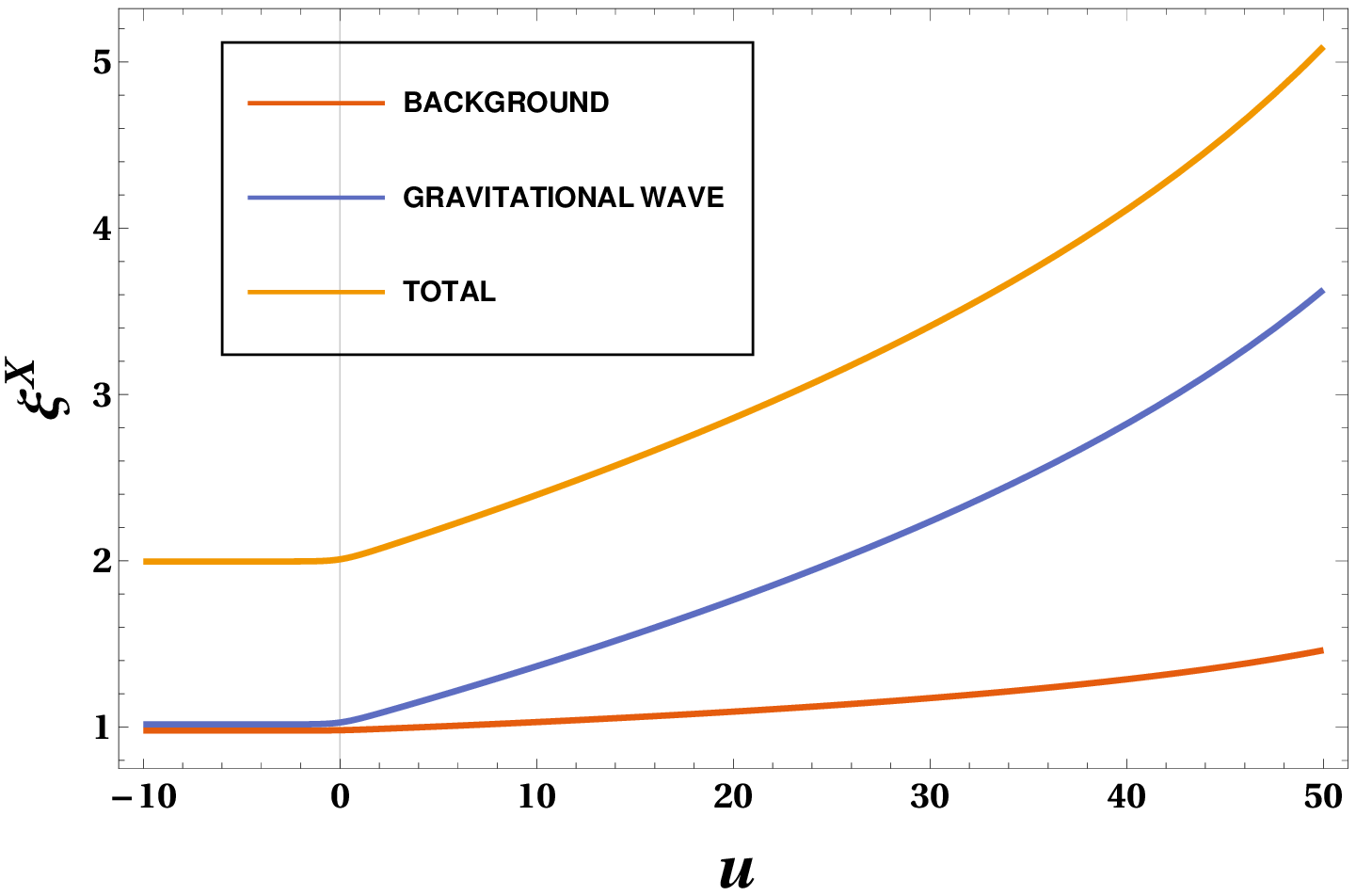}
		\caption{$\xi^X$ vs $u$}
		\label{fig:xi_X_+1}
\end{subfigure}\hspace{1cm}
	\begin{subfigure}[t]{0.45\textwidth}
		\centering
		\includegraphics[width=\textwidth]{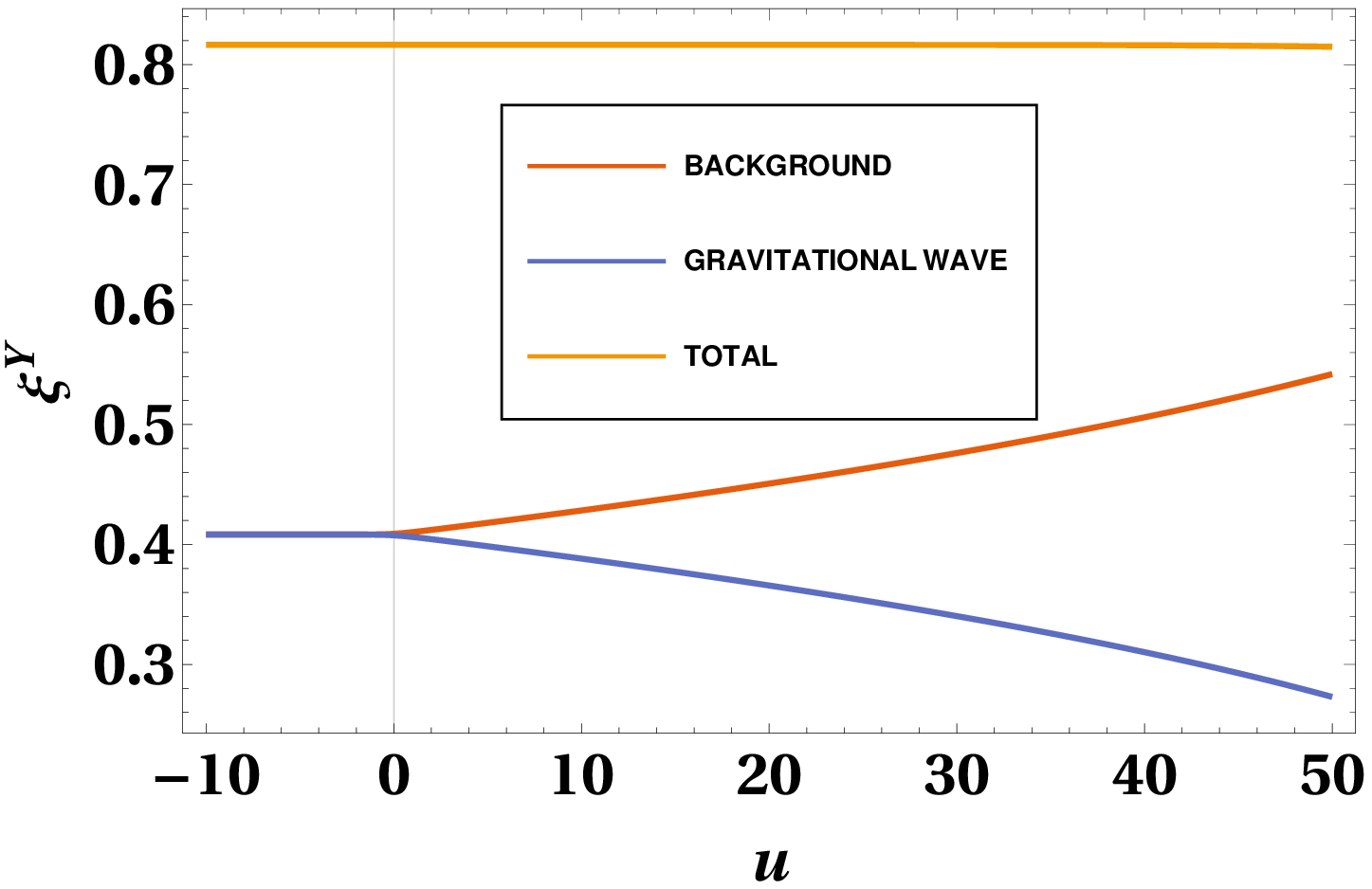}
		\caption{$\xi^Y$ vs $u$}
		\label{fig:xi_Y_+1}
	\end{subfigure}
	\caption{Plot of deviation vectors along the coordinates $X$ and $Y$ for $\omega=+1$.}
	\label{fig:deviation_+1_coo}
\end{figure}
\noindent The total deviation plot in Fig.(\ref{fig:xi_Y_+1}) shows it to be constant, a feature also obtained from the geodesics. In Fig.(\ref{fig:xi_X_+1}) for the deviation vector component along the $X$ direction, we find that the deviation for the gravitational wave part is more than that for the background.
The net deviation grows monotonically with the onset of the pulse. This result is consistent with the geodesic separation behaviour shown in Fig.(\ref{fig:x_1_geodesic}). The two analyses are not precisely equivalent due to the 
perturbative nature of geodesic deviation. Also,
we note that displacement memory is seen along both directions while frequency memory is not found.

\section{ Kundt metric with gyraton terms}
\noindent Having discussed the nature of memory effect for the Kundt wave geometry in Brans-Dicke theory, in this section we will incorporate the contribution from the gyratonic terms in the generalized Kundt metric. As, discussed in section II B, such a metric is generically given as:
\begin{equation*}
    ds^2=-Hdu^2-2dudv-2W_1dudx-2W_2dudy+\frac{1}{P^2}(dx^2+dy^2)
\end{equation*}
\noindent where, $P\equiv  P(u,x,y),\; H \equiv H(u,x,y)$ and the gyratonic terms $W_i \equiv W_i(u,x,y), $ $\forall \hspace{2mm} i  \hspace{2mm}\epsilon \hspace{2mm} \{x,y\}$.

\noindent Gyratonic spacetimes are ascribed to gravitational fields of spinning light beams \citep{Frolov:2005,Frolov1:2005}. The terms $W_i$ are responsible for angular momentum in the spacetime.  Gauge transformations can locally set $W_i$ to zero \citep{Podolsky:2014}. But, globally it is not possible to remove $W_i$ and, thus, the spacetime retains its rotational nature.

\noindent As before, we first solve the field equations to obtain the metric functions and the scalar field and thereafter, using them, we study the memory effect via the geodesic equations and also, geodesic deviation.

\noindent From the $G^u_{\;\;v}$ component of Eq.(\ref{eq:metric_var}) we have $\omega\;\phi_{,v}^2+\phi\;\phi_{,vv}=0$. We have considered a possible solution with $\phi,_v=0$, i.e. the scalar field is independent of the null coordinate $v$. Adding the $G^y_{\;\;x}$ and the $G^y_{\;\;y}$ components
of the Einstein tensor, we get:
\begin{eqnarray}
\phi_{,xx} + \phi_{,yy}=0 \label{eqn: Gux+Guy}
\end{eqnarray}
which is the same as Eq.(\ref{eq:phi_var}) in presence of a traceless matter field. In our case here, since we are considering a vacuum solution, this equation is consistent. 

\noindent From the $G^u_{\;\;u}$ component we get the following equation:
\begin{eqnarray}
-P^2 \Delta \log P=-\frac{P^2 \omega}{2\phi^2} \left(\phi_{,x}^2 +\phi_{,y}^2\right) \label{eqn: Guu} 
\end{eqnarray}

\noindent Similar to the previous analysis, in order to solve the above equations we decompose the metric functions and the scalar field as follows:

$\phi(u,x,y)=\alpha(u)\psi(x,y)$, \hspace{5mm} $P(u,x,y)=\dfrac{P'(x,y)}{U(u)}$, \hspace{5mm} $H(u,x,y)=H'(u)h(x,y)$. 

\noindent Using this decomposition in Eq.(\ref{eqn: Gux+Guy}) we get:
\begin{eqnarray}
\psi(x,y)= a+\log[x^2 + y^2] \label{eqn: psi}
\end{eqnarray}

\noindent Substituting the new functional forms in Eq.(\ref{eqn: Guu}) we obtain:
\begin{eqnarray}
P'(x,y)= \frac{\sqrt{x^2+y^2}}{(a+\log[x^2+y^2])^{\frac{\omega}{2}}}
\end{eqnarray}

\noindent The equation that follows from the $G^x_{\;\;u}$ component is:
\begin{eqnarray}
\left[\left(\frac{P_{,u}}{P}\right)_{,x}+PP_{,y}(W_{1,y}-W_{2,x}) + \frac{P^2}{2}(W_{1,yy}-W_{2,xy})\right]= \nonumber \\ \left[(\phi_{,x}\phi_{,u})\frac{\omega}{\phi^2} + \frac{\phi_{,xx}}{\phi} + \frac{P_{,u}}{P} \frac{\phi_{,x}}{\phi} -\frac{P^2}{2} \frac{\phi_{,y}}{\phi}(W_{1,y}-W_{2,x})\right] \label{eqn: Gxu}
\end{eqnarray}

\noindent To solve the above equation we decompose the cross terms as follows:
\begin{eqnarray}
W_1(u,x,y)=-\frac{J(u) y}{2[x^2+y^2]}\;\;\;\; W_2(u,x,y)=\frac{J(u) x}{2[x^2+y^2]} \label{decom: W1 W2}
\end{eqnarray}
where $J(u)$ is some function of $u$. We will see, while solving the geodesic equations that $J(u)$ may be associated with angular momentum. 

\noindent From Eq.(\ref{decom: W1 W2}) we note that
\begin{eqnarray}
W_{1,y}-W_{2,x}=0 
\end{eqnarray}
Using the above result in Eq.(\ref{eqn: Gxu}) we find:
\begin{eqnarray}
\frac{\alpha_{,u}}{\alpha} \frac{\psi_{,x}}{\psi}(\omega+1) - \frac{U_{,u}}{U}\frac{\psi_{,x}}{\psi}=0 \nonumber 
\end{eqnarray}
Thus, $\alpha(u)$ and $U(u)$ satisfy:
\begin{eqnarray}
\frac{\alpha_{,u}}{\alpha}(\omega+1)= \frac{U_{,u}}{U}
\end{eqnarray}
This is the same as the Eq.(\ref{eq:relation}) in the Kundt wave metric case.
The last equation that will constrain the metric function is from the $G_{uu}$ component. It leads to,
\begin{eqnarray}
\frac{1}{8P^2}[4P^4(H_{,xx} +H_{,yy}) + 8P^3H(P_{,xx}+P_{,yy}) - 8P^2H(P_{,x}^2 +P_{,y}^2) + 16P_{,uu}P -32P_{,u}^2]= \nonumber \\
\left[\phi_{,u}^2+ \frac{H}{2}P^2(\phi_{,x}^2 + \phi_{,y}^2)\right]\frac{\omega}{\phi^2} + \frac{1}{\phi}\left[\phi_{,uu}-\frac{P^2}{2}(-2W_{1,u}+H_{,x})\phi_{,x} -\frac{P^2}{2}(-2W_{2,u}+H_{,y})\phi_{,y}\right] \label{eqn: Gdownuu}
\end{eqnarray}
From both the Eqs.(\ref{eqn: psi}) and (\ref{decom: W1 W2}) one can see that:
\begin{eqnarray}
W_{1,u}\phi_{,x} + W_{2,u}\phi_{,y}=0 
\end{eqnarray}

\noindent As in Kundt metric case we solve for $U(u)$=constant. This reduces Eq.(\ref{eqn: Gdownuu}) to:
\begin{eqnarray}
h_{,xx}+ h_{,yy} + \left[h_{,x} \frac{\phi_{,x}}{\phi} + h_{,y} \frac{\phi_{,y}}{\phi}\right]=0 \nonumber
\end{eqnarray}
The solution is the same as for the Kundt wave case.
\begin{eqnarray}
H(u,x,y)=H'(u) \log[a+\log[x^2 + y^2]]
\end{eqnarray}
Thus, on solving the field equations we have obtained all the metric functions except $H'(u)$ and $J(u)$. These functions are unconstrained and are the source of the gravitational wave and the gyratonic contribution respectively. We will choose these functions judiciously while solving for the geodesics and the geodesic deviation.

\noindent Following our approach for Kundt waves,  we rewrite the gyratonic metric in new coordinates defined via the coordinate transformations: $x=e^{X-a/2}\cos Y$ and $y=e^{X-a/2}\sin Y $. As earlier, we find that the scalar $\psi$ (realted to the $\phi$) is twice of $X$. 

\noindent The metric line element in coordinates (u,v,X,Y) is given as: 
\begin{eqnarray}
ds^2=-H'(u) \log[2X]du^2 - 2dudv - J(u)dudY + (2X)^\omega (dX^2 + dY^2) \label{eqn: metric GKundt}
\end{eqnarray}
 We can see that on comparing with the Kundt wave metric, the only difference is the presence of the $J(u)dudY$ term in the metric. 
 We will see that this term will have its contribution 
 to the overall memory effect.

\subsection{Displacement memory from geodesic analysis}
\noindent The geodesic equations corresponding to $X$ and $Y$, for the metric line element given by Eq.(\ref{eqn: metric GKundt}) are as follows.
\begin{eqnarray}
\frac{d^2X}{du^2} + \frac{\omega}{2X}\left[\left(\frac{dX}{du}\right )^2  - \left(\frac{dY}{du}\right)^2 \right]
+ \frac{H'(u)}{(2X)^{\omega+1}} =0 \label{eqn: geodesic GKundt X}\\ 
\frac{d^2Y}{du^2} + \frac{\omega}{X}\frac{dX}{du}\frac{dY}{du} - \frac{1}{2(2X)^\omega} \frac{dJ(u)}{du}=0 \label{eqn: geodesic GKundt Y}
\end{eqnarray}

\noindent We find that $\ddot{u}=0$  from the geodesic equation of $v$. Therefore $u$ may be chosen as as an affine parameter. From Eq.(\ref{eqn: geodesic GKundt Y}), it is clear that the geodesic along $Y$ coordinate is dependent on $J(u)$. This is different from the Kundt wave scenario where along $Y$ there could be no evolution.  Hence, Eq.(\ref{eqn: geodesic GKundt X}) is dependent on both $J(u)$ and $H'(u)$. 
It is evident that the presence of cross terms involving the 
gyratonic contribution will have a significant impact on the evolution of the geodesics.

\noindent The displacement memory effect for coordinate $X$ obtained below is equivalent to the memory due to the scalar field, as justified earlier. 

\noindent Proceeding further, we note that the first integral of Eq.(\ref{eqn: geodesic GKundt Y}) gives, 
\begin{equation}
    \dot{Y}=\frac{C+J(u)}{2(2X)^\omega}\label{eq:1st_integral_Y}.
\end{equation}
\noindent Here, $C$ is a constant of integration which can be set to zero by noting that 
$\dot{Y}$ vanishes where $J(u)=0$. Since $X$ is functionally related to $\psi[X(u)]$, we infer from Eq.(\ref{eq:1st_integral_Y}) that the displacement memory along coordinate $Y$ is both due to scalar field and the gyraton.
\noindent As in the case without the gyraton term, we can, for
$\omega\neq -2$, convert the equation for X to that of a
generalised Levinson-Smith system  with the added feature that
function $g(x)$ in (32), is now $g(x,t)$. We obtain,
\begin{equation}
    \frac{d^2q}{du^2}-\frac{1}{2q}\bigg(\frac{dq}{du}\bigg)^2+
    - \frac{\omega(\omega+2)}{4 q^{\frac{\omega}{\omega+2}}} J^2 (u) + 2(\omega+2)H'(u)=0 \label{eq:levinson_smith1}
\end{equation}

\noindent Asymptotic behaviour of Eq.(\ref{eq:levinson_smith1}) has been studied in \citep{TUNC:2007}. Such solutions are bounded and converge as $u\to +\infty$.

\noindent Let us now separately deal with the constant curvature solution
($\omega=-2)$ and singular solutions ($\omega\neq-2$). The Ricci scalar in each case turns out to be the same as for the Kundt wave metric.
We assume $H'(u)=\frac{\sech^2(u)}{2}$ (same as in the Kundt wave metric) and $J(u)=b \sech^2(u)$. \footnote{We will set the parameter $b=1$ for 
our analysis.} Thus, both the wave and gyratonic terms are dominant for a 
limited and finite duration of "time"({\em i.e.} $u$).

\begin{itemize}
    \item \underline{$\omega=-2$}
    \newline This case corresponds to $R=-8$, a constant negative curvature solution. We solve the geodesic equations (\ref{eqn: geodesic GKundt X}) and (\ref{eqn: geodesic GKundt Y}) for this $\omega$ numerically in {\em Mathematica 10} and plot the solutions.
    
    \begin{figure}[H]
	\centering
	\begin{subfigure}[t]{0.4\textwidth}
		\centering
		\includegraphics[width=\textwidth]{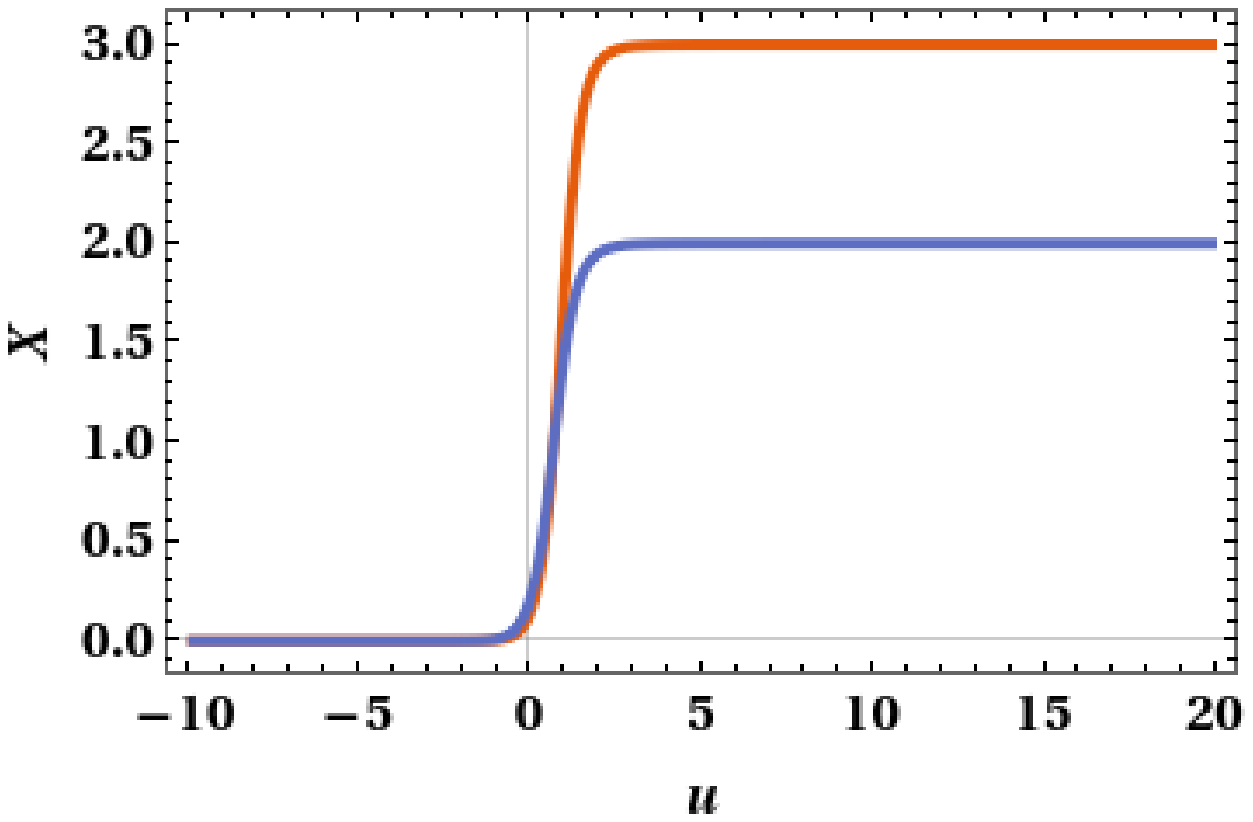}
		\caption{\small \centering 
		}
		\label{fig:Gk_X_-2_geodesic}
		\end{subfigure} \hspace{1.5cm}
	\begin{subfigure}[t]{0.4\textwidth}
		\centering
		\includegraphics[width=\textwidth]{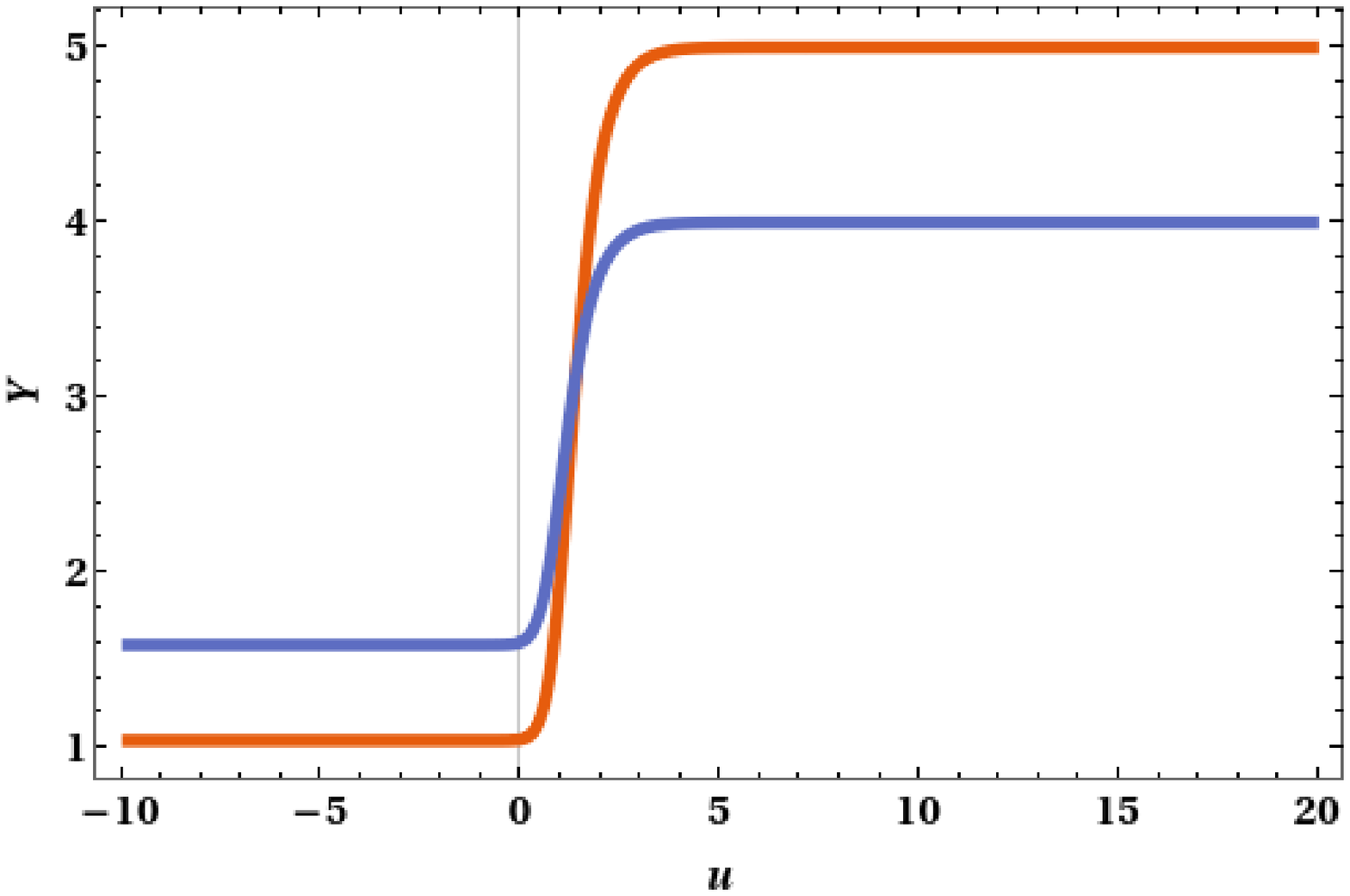}
		\caption{\small \centering
		}
		\label{fig:Gk_Y_-2_geodesic}	
		\end{subfigure}
		\caption{Geodesics for gyratonic Kundt metric with $\omega=-2$.}
		\label{fig: geo_gyr_omega_-2}
		\end{figure}
		
		 \noindent The geodesic evolution show a constant shift
		 along $u$ as well as a displacement memory effect along both $X$ and $Y$ directions.  Apart from the shift along $u$ (which is present in
		 the evolution of both $X$ and $Y$), Fig.(\ref{fig:Gk_Y_-2_geodesic}), i.e. the evolution of $Y$, also 
		 shows the role the gyratonic terms in the metric. The plot along the $X$ direction in Fig.(\ref{fig:Gk_X_-2_geodesic}) is similar 
		 (but shifted along $u$) to the Kundt wave case.
		 
	\item \underline{$\omega\neq-2$}
	\newline Since the Ricci scalar is the same as for 
	the Kundt wave metric, we find that for $\omega\neq2$ there is a singularity at $X=0$. We have carefully restricted our geodesic analysis in the domain $X>0$. The plots for the geodesics are shown below.
	
	\begin{figure}[H]
	\centering
	\begin{subfigure}[t]{0.4\textwidth}
	\centering
		\includegraphics[width=\textwidth]{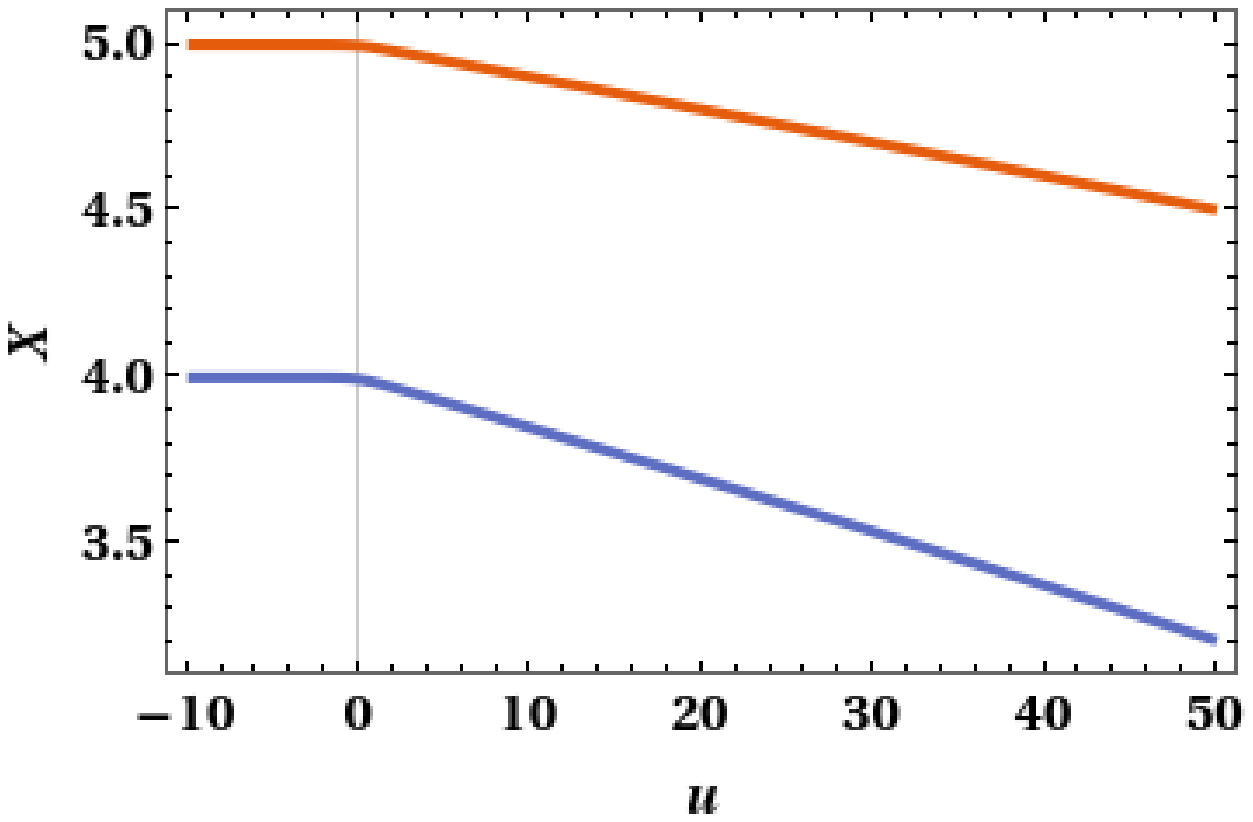}
		\caption{\small Initial positions of $X$ are 5 (orange) and 4 (blue) respectively.}
		\label{fig:Gk_X_1_geodesic}
		\end{subfigure} \hspace{1cm}
	\begin{subfigure}[t]{0.4\textwidth}
		\centering
		\includegraphics[width=\textwidth]{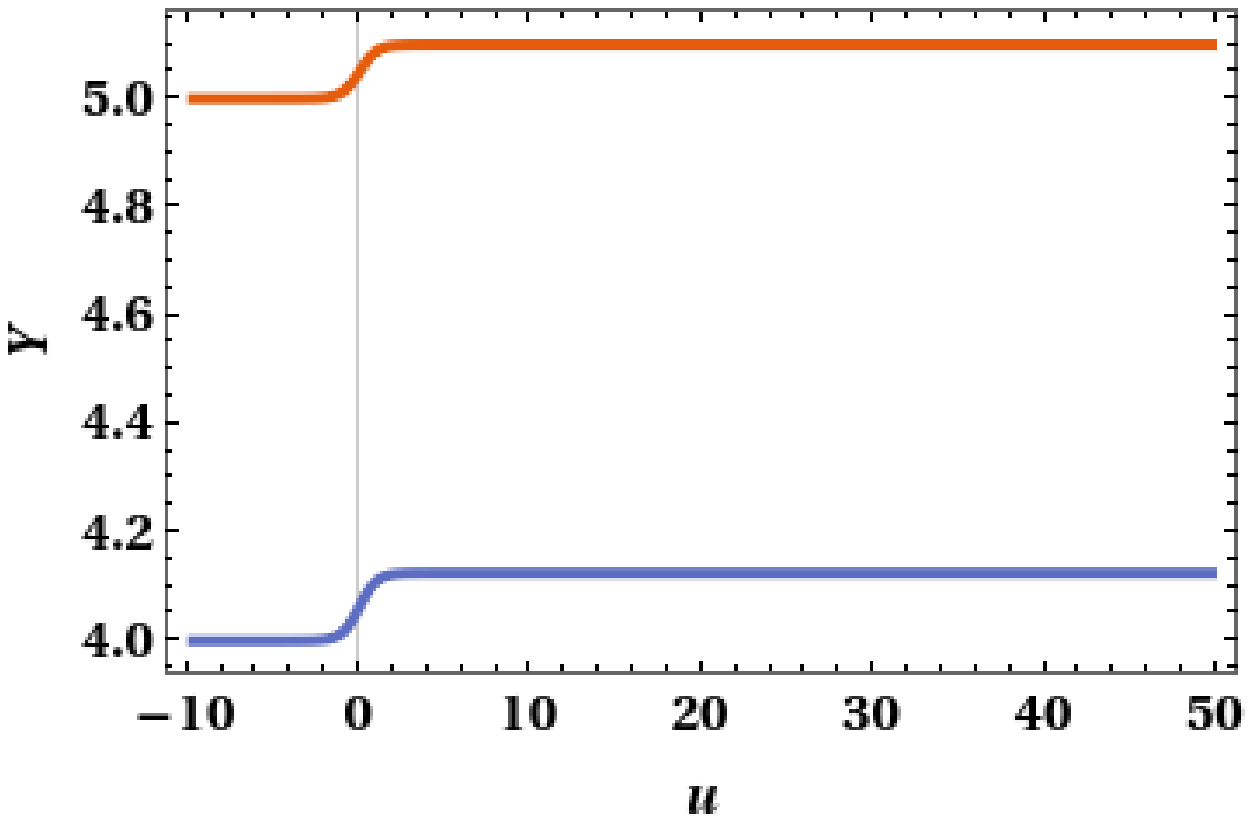}
		\caption{\small  Initial positions of $Y$ are 5 (orange) and 4 (blue) respectively.}
		\label{fig:Gk_Y_1_geodesic}	
		\end{subfigure}
		
		\caption{Geodesics for gyratonic Kundt metric with $\omega=+1$.}
		\label{fig:geo_gyr_omega_+1}
		\end{figure}
		
From Fig.(\ref{fig:geo_gyr_omega_+1}) we see that the separation slowly builds up along the $X$ direction. We note from Eq.(\ref{eq:1st_integral_Y}) that $\dot{Y}=J(u)/(4X)$.  Since $J(u)$ is a sech-squared pulse, we find that $\dot{Y}$ is zero at $u\to \pm \infty$. The only significant contribution of the gyratonic pulse is centered at $u=0$. Thus, for the gyratonic Kundt metric, both the $X$ and $Y$ coordinates have memory effect around $u=0$ in contrast to memory only along  the $X$ direction for Kundt waves.   

\end{itemize}

\noindent From the plots of the geodesics it is clear that the most visible contribution of the gyratonic term is in the evolution along  for $Y$. 
As for the evolution along  X, the plots are shifted because of the additional term dependent on $\dot{Y}^2\propto J[u]^2$ in the equation (\ref{eqn: geodesic GKundt X}).

\subsection{Geodesic deviation analysis of the memory effect}

\noindent The orthonormal tetrads for the gyratonic Kundt metric (\ref{eqn: metric GKundt}) are written below.
\begin{equation}
\begin{split}
& e_0\,^{\mu}=[1,\dot{v},\dot{X},\dot{Y}]\hspace{2.5cm}
 e_3\,^{\mu}=[-1,1-\dot{v},-\dot{X},-\dot{Y}]\\
 & e_1\,^{\mu}=[0,-(2X)^{\omega/2}\dot{X},-(2X)^{-\omega/2},0] \\
  & e_2\,^{\mu}=[0,-(2X)^{\omega/2}\dot{Y}+(J/2)(2X)^{-\omega/2},0,-(2X)^{-\omega/2}]   \label{eq:tetrad_withc}
    \end{split}
\end{equation}

\begin{figure}[H]
    \centering
	\begin{subfigure}{.8\textwidth}
		\centering
		\includegraphics[width=\textwidth]{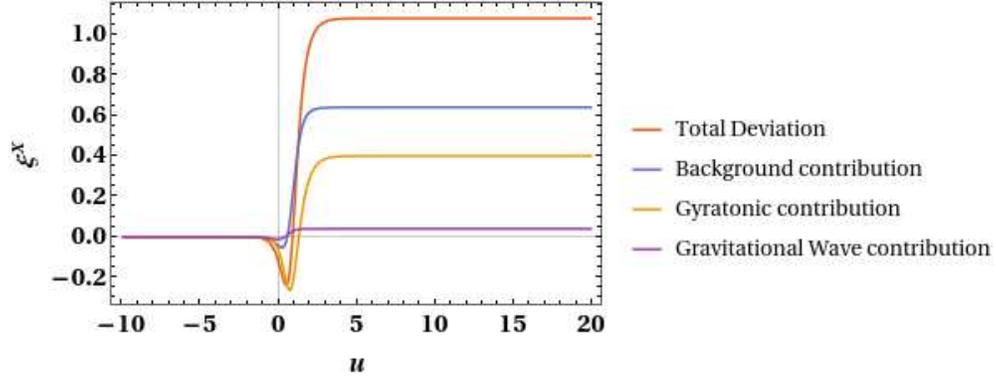}
		\caption{\small Deviation along $X$ .}
		\label{fig:deviation GKundt_w=-2_ZX}
		\end{subfigure} \vspace{-1cm}
    \begin{subfigure}{.8\textwidth}
		\centering
		\includegraphics[width=\textwidth]{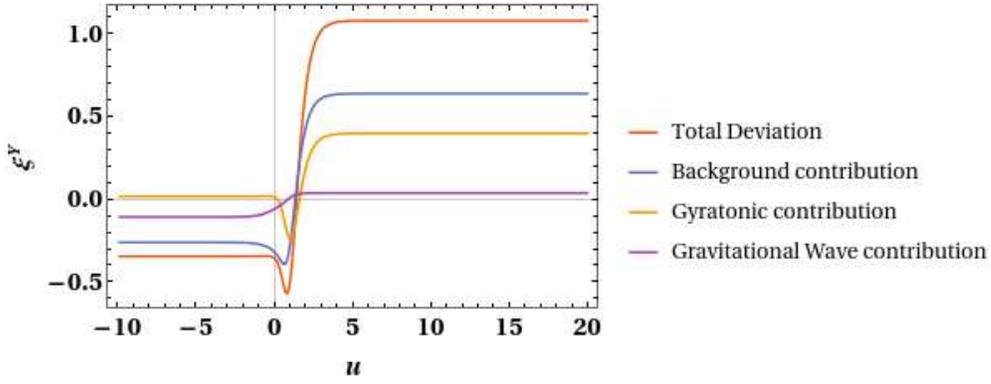}
		\caption{\small Deviation along $Y$.}
		\label{fig:deviation GKundt_w=-2_ZY}
		\end{subfigure} \hspace{1cm}\vspace{1cm}
	\caption{Deviation plots for $\omega=-2$}
    \label{fig: deviation GK w=-2}
\end{figure}

\begin{figure}[H]
    \centering
    \begin{subfigure}{.4\textwidth}
     \centering
		\includegraphics[width=\textwidth]{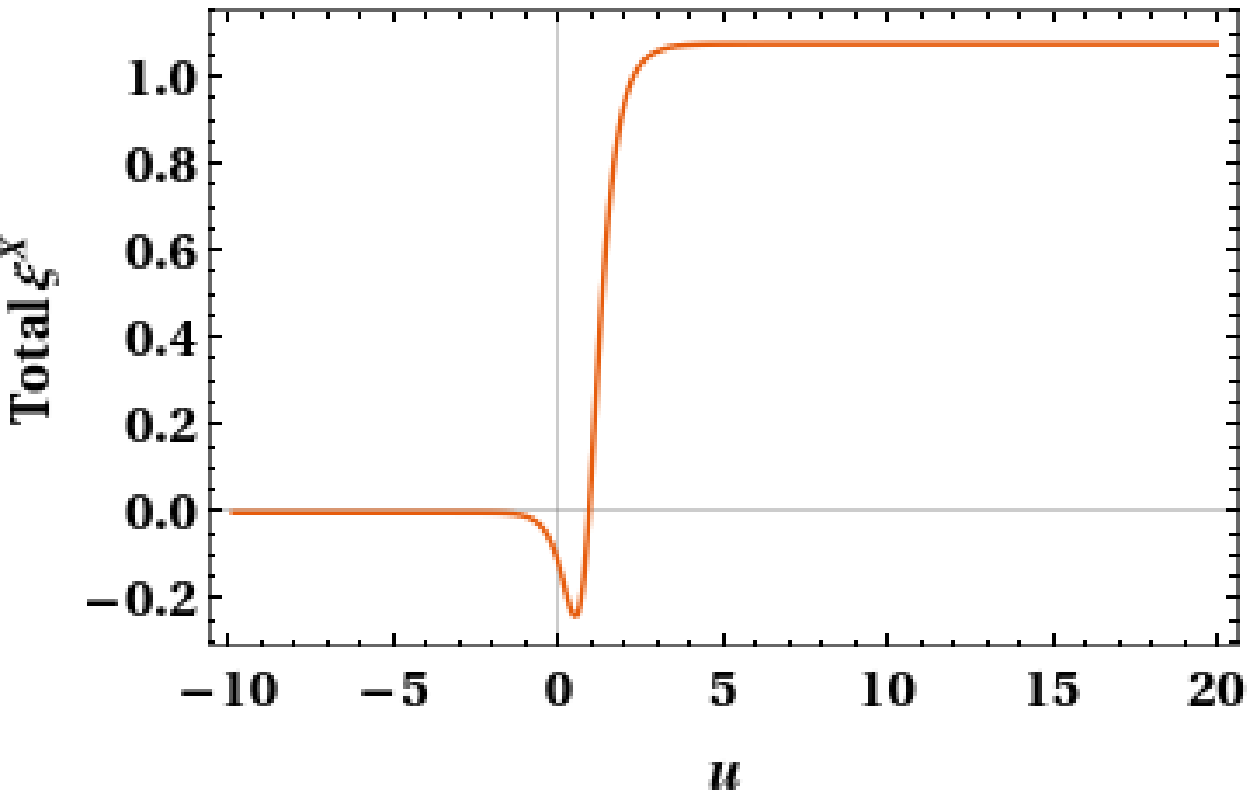}
		\caption{\small  Total deviation along $X$}
		\label{fig:deviation GKundt_w=-2_ZX_total}	
	\end{subfigure}\hspace{1cm}
       \begin{subfigure}{.4\textwidth}
		\centering
		\includegraphics[width=\textwidth]{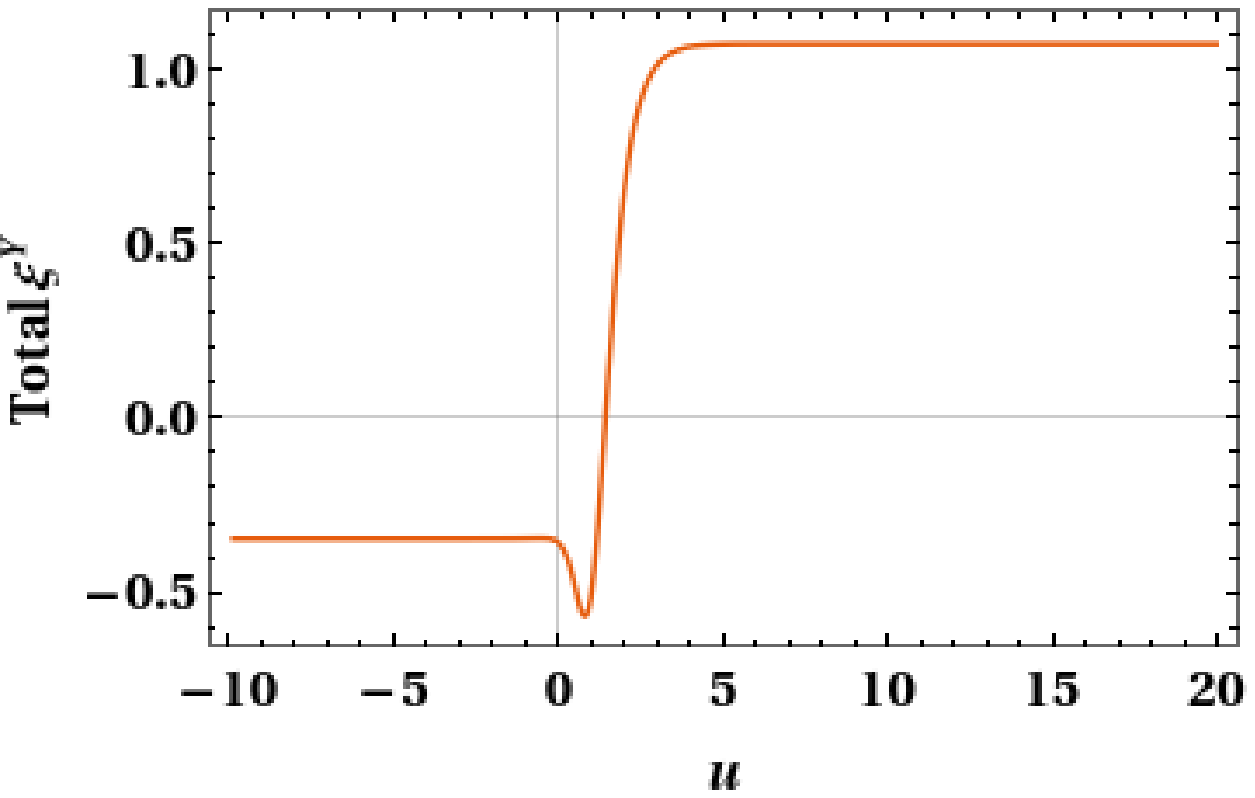}
		\caption{\small Total deviation along $Y$.}
		\label{fig:deviation GKundt_w=-2_ZY_total}	
		\end{subfigure}
    \caption{Total deviation plots for $\omega=-2$}
    \label{fig: tdeviation GK w=-2}
\end{figure}

\begin{figure}[H]
    \centering
	\begin{subfigure}{.8\textwidth}
		\centering
		\includegraphics[width=\textwidth]{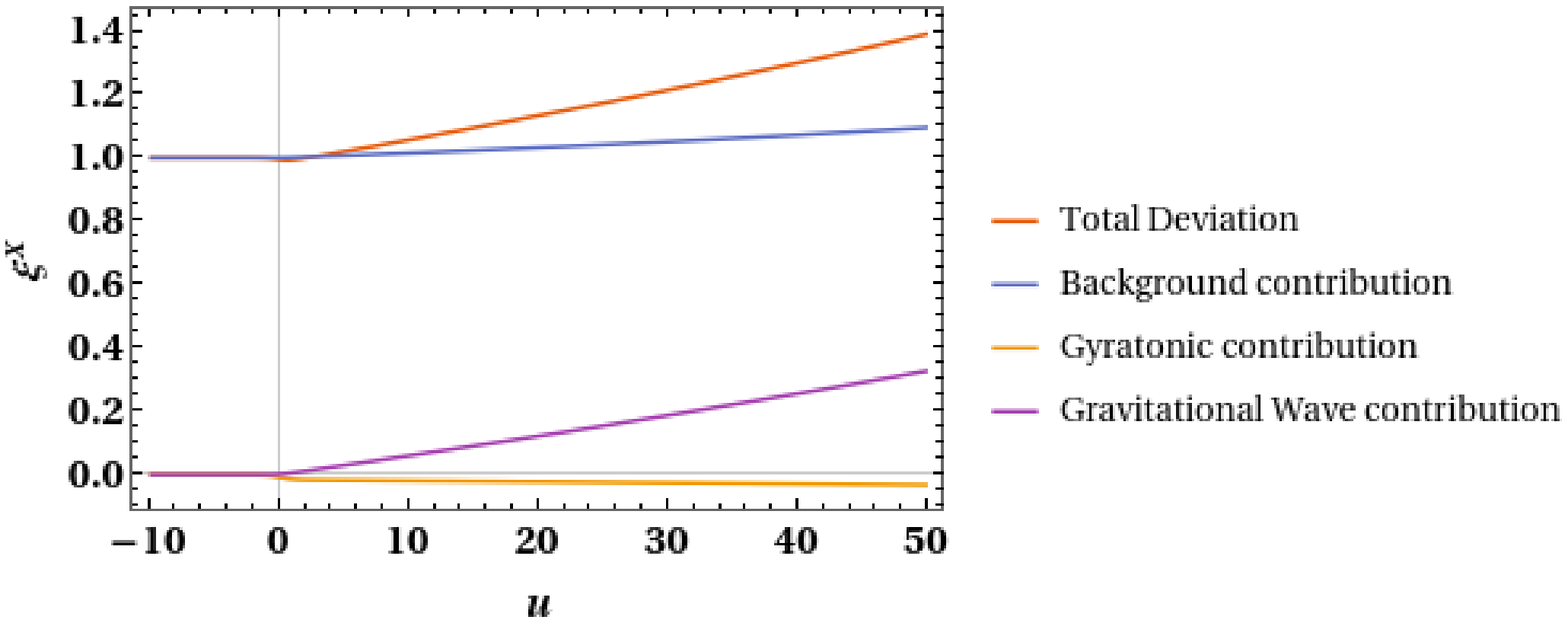}
		\caption{\small Deviation along $X$ .}
		\label{fig:deviation GKundt_w=1_ZX}
		\end{subfigure} \hspace{1cm} \vspace{1cm}

    \begin{subfigure}{.8\textwidth}
		\centering
		\includegraphics[width=\textwidth]{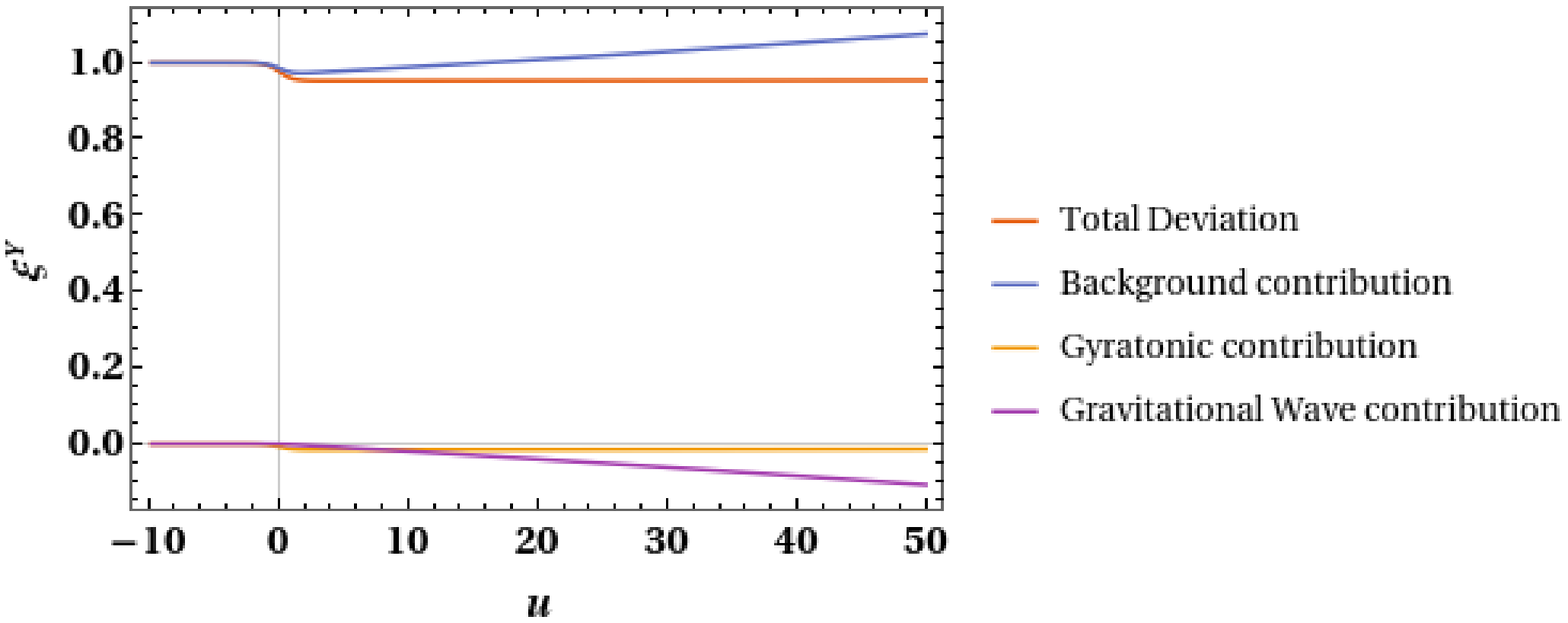}
		\caption{\small Deviation along $Y$.}
		\label{fig:deviation GKundt_w=1_ZY}
		\end{subfigure} \hspace{1cm}\vspace{1cm}
		\caption{Deviation Plots for $\omega=1$}
    \label{fig: deviation GK w=1}
\end{figure}

\begin{figure}[H]
    \centering
    \begin{subfigure}{.4\textwidth}
     \centering
		\includegraphics[width=\textwidth]{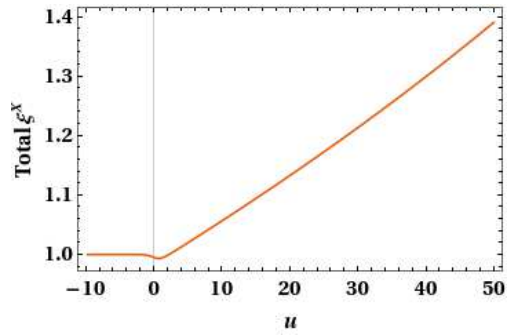}
		\caption{\small  Total deviation along $X$}
		\label{fig:deviation GKundt_w=1_ZX_total}	
	\end{subfigure}\hspace{1cm}
       \begin{subfigure}{.4\textwidth}
		\centering
		\includegraphics[width=\textwidth]{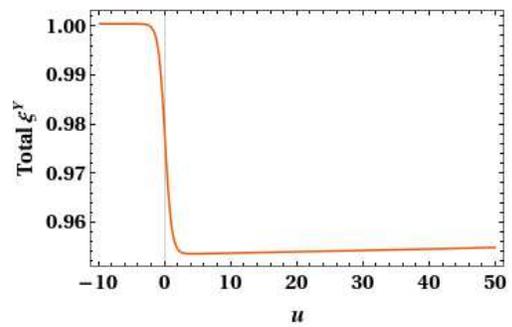}
		\caption{\small Total deviation along $Y$.}
		\label{fig:deviation GKundt_w=1_ZY_total}	
		\end{subfigure}
    \caption{Total deviation plots for $\omega=1$}
    \label{fig: tdeviation GK w=1}
\end{figure}

\noindent The gyratonic term $J(u)$ is only present in  $e_2\,^\mu$. We now check whether the constructed tetrads are parallely transported, i.e.
($U^\mu=e_0^{\;\;\mu}$).
\begin{eqnarray}
U\cdot D e_0^{\;\;\mu}=0,\;\;
U\cdot D e_1^{\;\;\mu}= \frac{\omega \dot{Y}}{2X}e_2^{\;\;\mu},\;\;
U\cdot D e_2^{\;\;\mu}=-\frac{\omega\dot{Y}}{2X}e_1^{\;\;\mu},\;\;
U\cdot D e_3^{\;\;\mu}=0
\end{eqnarray}
Therefore, if we rotate $e_1^{\;\;\mu}$ and $e_2^{\;\;\mu}$ by an angle $\theta_p$, such that it satisfies $\dot{\theta_p}=\frac{\omega \dot{Y}}{2X}$, we will obtain a parallely propagating tetrad. Thus, $\theta_p$ is dependent on the gyratonic term $J(u)$, following from Eq.(\ref{eq:1st_integral_Y}). In the new tetrad basis, we calculate the nonvanishing Riemann components and separate them into background, gyratonic and the gravitational wave parts. The deviation equation for the three 
parts are
\begin{gather}
    \frac{d^2Z^i_B}{du^2}=-({R}^i\,_{0j0})_B Z^j_B \label{eq:deviation_bg_gyraton}\\
    \frac{d^2Z^i_G}{du^2}=-[({R}^i\,_{0j0})_B+({R}^i\,_{0j0})_G] Z^j_G-({R}^i\,_{0j0})_G Z^j_B\label{eq:deviation_gyraton_gyraton}\\
    \frac{d^2Z^i_W}{du^2}=-[({R}^i\,_{0j0})_B+({R}^i\,_{0j0})_G+({R}^i\,_{0j0})_W]Z^j_W-({R}^i\,_{0j0})_W(Z^j_B+Z^j_G).\label{eq:deviation_wave_gyraton}
\end{gather}
\noindent The deviation equations for the gyratonic Kundt metric 
are straightforward generalizations of Eqs.(\ref{eq:deviation_eqn1_bg}) and (\ref{eq:deviation_eqn1_wave}) where there was no contribution coming from the term $(R^i\,_{0j0})_G$.  The expressions for the Riemann tensor components in the tetrad frame for these three components are given in the Appendix [see Eqs.(\ref{eq:R110B})-(\ref{eq:R2020W})]. The  geodesic deviation equations (\ref{eq:deviation_bg_gyraton}), (\ref{eq:deviation_gyraton_gyraton}) and (\ref{eq:deviation_wave_gyraton}) were numerically solved in {\em Mathematica 10} and the deviation vectors were obtained first in the tetrad basis. Specific initial values are assumed for each part of the
deviation vector (i.e. background, wave and gyraton) 
at a $u$ value reasonably far from where the pulse $H'(u)$
or the gyratonic term $J(u)$ is significant (i.e. near $u=0$).
Thereafter,  following the procedure adopted for the 
Kundt waves, the deviation vectors are transformed back to the coordinate basis 
where the evolution is analysed. 
The plots thus generated appear in Figs.(\ref{fig: deviation GK w=-2}), (\ref{fig: tdeviation GK w=-2}) 
and Figs.(\ref{fig: deviation GK w=1}) and (\ref{fig: tdeviation GK w=1})
for $\omega=-2$ and $\omega=1$ respectively. 

\noindent The evolution of each part of the deviation vector (i.e. background, wave and
gyraton) as shown explicitly in  
Fig.(\ref{fig: deviation GK w=-2}) and Fig. (\ref{fig: deviation GK w=1}),
demonstrate their individual contributions. The qualitative similarity of the total deviation plots [Fig. (\ref{fig: tdeviation GK w=-2}) and Fig. (\ref{fig: tdeviation GK w=1})]
with the results found using geodesics [Fig. (\ref{fig: geo_gyr_omega_-2}) and Fig. (\ref{fig:geo_gyr_omega_+1})] is quite evident. On comparing the plots for deviation in Kundt wave metric [Fig. (\ref{fig:deviation_-2_coo}) and Fig. (\ref{fig:deviation_+1_coo})] with deviation plots for the
gyratonic Kundt metric we note that along the $X$ direction, the plots are very similar, except for the additional gyratonic contribution. However, along the $Y$ direction the plots are significantly different. In the 
Kundt wave case, background and gravitational wave contributions cancel each other, whereas in the gyratonic Kundt metric, the contribution sums up and gives an effective displacement memory.

\section{Conclusions}

\noindent  Let us now summarise pointwise, the main results obtained in this article.

\noindent $\bullet$ In vacuum Brans--Dicke theory we
have constructed solutions representing a Kundt-type
line element without and with gyratonic terms. The solutions have the expected feature of two unspecified functions $H'(u)$ and $J(u)$ which represent the wave-profile and the gyraton term respectively. Both these functions can be specified
while writing down explicit solutions. We choose them to be localised fucntions (eg. proportional to $sech^2 u$) with the aim of studying memory effects.

\noindent $\bullet$ For both line elements with
chosen profiles for the $H'(u)$ and $J(u)$ we first
study geodesics and then geodesic deviation in order to obtain memory effects. Logical similarities between results using geodesics and geodesic deviation are visible in our results. Displacement memory via geodesics and memory found using 
solutions of the deviation equation explicitly demonstrate the dependencies on the presence and 
profiles of the functions $H'$ and $J$. The role of the gyraton term is clearly visible in the memory effects.

\noindent $\bullet$ Displacement memory for $\omega=-2$ and without the gyraton term is exactly
solvable and easy to understand analytically--a feature we show while analysing this specific case
with constant negative Ricci scalar.

\noindent $\bullet$ In our analysis of deviation
we have used a parallely transported Fermi basis 
where the equations simplify. We have split the
deviation vector into background, gravitational wave
and gyratonic parts and have shown how each part
influences the behaviour of
the total deviation vector, thereby ensuring
the existence of memory. 

\noindent $\bullet$  The identification of the main geodesic equation for general
$\omega$ with a known
dynamical system, the Levinson-Smith system, is an interesting observation in our work.
The lack of exact solutions of such systems was a hindrance in using this correspondence
for understanding memory. As mentioned earlier, it may be possible to 
extend and enrich our work along these lines using inputs from mathematics and the 
theory of dynamical systems. 

\

\noindent The memory effects shown here for
the two different Kundt metrics are characteristic
features of BD theory itself. This is apparent
through the dependencies of memory on $\omega$ and/or the
BD scalar field. Thus, there is an intrinsic difference with similar scenarios in GR at 
a qualitative as well as a quantitative level.

\noindent The geodesic deviation analysis
carried out here (using the Fermi basis etc.) can readily be applied to calculate memory effects for any spacetime, in particular those where radiative
behaviour is present. Further, one may also
employ the ${\cal B}$-matrix formalism as introduced in \citep{Loughlin:2019} to study the behaviour of the kinematic variables of timelike geodesic congruences. One may also search for memories in impulsive gravitational wave spacetimes \citep{Zhang:2018,Srijit:2019} and compare with the results obtained here, by setting appropriate limits.

\noindent Finally, it may be worthwhile 
identifying Kundt geometries in other alternative
theories so that a comparison can be made on the
nature and distinguishing features of the memory effects manifest in each such theory. Even within BD theory, there  exists scope of finding newer Kundt spacetimes for which memory effects can always be
investigated using the methods outlined here.
We hope to return to these issues in future work.

\section*{ACKNOWLEDGEMENTS}

\noindent SS thanks the Department of Physics, IIT Kharagpur, India for providing him with the opportunity to work on this project during his tenure as an integrated M.Sc. student. IC acknowledges Matthias Blau for discussions. IC also thanks the University Grants Commission (UGC), Government of India for  providing  financial  assistance  through  a senior research fellowship (SRF) (reference ID: 523711).  


\bibliographystyle{apsrev4-2}
\bibliography{mybibliography_bd}

\section*{APPENDIX}

\subsection*{Einstein tensor components}

\noindent The Einstein tensor components of the generalized Kundt metric with gyratons are given below.
\begin{flalign}
&G^u\,_{v}= 
G^x\,_{x} =
G^y\,_{y}=0 \\
&G^u\,_{u}=-P^2 \Delta \log[P] \\
&G^x\,_{u}= \left[\left(\frac{P_{,u}}{P}\right)_{,x}+PP_{,y}(W_{1,y}-W_{2,x}) + \frac{P^2}{2}(W_{1,yy}-W_{2,xy})\right] \\
&G^y\,_u=\left[\left(\frac{P_{,u}}{P}\right)_{,y}+PP_{,y}(W_{2,x}-W_{1,y}) + \frac{P^2}{2}(W_{2,xx}-W_{1,xy})\right] \\
&G_{uu}=\frac{1}{8P^2}[4P^4(H_{,xx} +H_{,yy}) + 8P^3H(P_{,xx}+P_{,yy}) - 8P^2H(P_{,x}^2 +P_{,y}^2) + 16P_{,uu}P -32P_{,u}^2] 
\end{flalign}

\noindent Note that by setting $W_1=W_2=0$ we get the Einstein tensors for the Kundt wave metric given in Eq.(\ref{eq:metric_Kundt_1}).

\subsection*{Riemann tensor in tetrad frame for Kundt spacetimes with gyraton terms}

\noindent The Riemann tensor in the parallely propagated tetrad frame for the gyratonic Kundt metric are as follows.

\noindent {\em Background}
\begin{flalign}
    &(R^1_{\;\;010})_B=\frac{\omega}{2X^2}[\sin (\theta_p)\dot{X}+\cos(\theta_p)\dot{Y}]^2 \label{eq:R110B}\\
    &(R^1_{\;\;020})_B=(R^2_{\;\;010})_B=-\frac{\omega}{4X^2}[\sin(2\theta_p)(\dot{X}^2-\dot{Y}^2)+2\cos(2\theta_p)\dot{X}\dot{Y}]\\
    &(R^2_{\;\;020})_B=\frac{\omega}{2X^2}[\cos(\theta_p)\dot{X}-\sin (\theta_p)\dot{Y}]^2
\end{flalign}
\noindent {\em Gyraton}
\begin{flalign}
   &(R^1_{\;\;010})_G=-\frac{\omega X}{(2X)^{\omega+2}}\dot{J}(u)\sin (2\theta_p)\\
&(R^1_{\;\;020})_G=(R^2_{\;\;010})_G=\frac{\omega X}{(2X)^{\omega+2}}\dot{J}(u)\cos(2\theta_p)\\
     &(R^2_{\;\;020})_G=\frac{\omega X}{(2X)^{\omega+2}}\dot{J}(u)\sin (2\theta_p)
\end{flalign}
\noindent {\em Gravitational wave}
\begin{flalign}
    &(R^1_{\;\;010})_W=-\frac{H'(u)}{(2X)^{\omega+2}}[1+(\omega+1)\cos (2\theta_p)]\\
     &(R^1_{\;\;020})_W=(R^2_{\;\;010})_W=-\frac{\omega+1} {(2X)^{\omega+2}}H'(u)\sin(2\theta_p)\\
      &(R^2_{\;\;020})_W=\frac{H'(u)}{(2X)^{\omega+2}}[(\omega+1)\cos (2\theta_p)-1]\label{eq:R2020W}
\end{flalign}

\noindent  An overdot denotes differentiation w.r.t. $u$.

\end{document}